\documentclass[useAMS,usenatbib]{mnras}

\usepackage[pdftex]{graphicx}
\usepackage{color}
\usepackage{amsmath}  
\usepackage{amssymb}  
\usepackage{upgreek}  
\usepackage{afterpage}  
\usepackage{times}  
\usepackage{paralist}  
\usepackage{pbox}  
\usepackage{booktabs}  
\usepackage[flushleft]{threeparttable}  
\usepackage{mathtools}  
\usepackage{physics}
\usepackage{siunitx}
\usepackage{multirow}
\usepackage{graphicx}
\usepackage{subfig}
\usepackage{float}
\usepackage{csquotes}
\newcommand{\qtty}{\quantity}
\usepackage[nameinlink]{cleveref}
\crefname{figure}{Fig.}{Figures}
\crefname{table}{Table}{Tables}
\crefname{section}{Section}{Sections}
\crefname{subsection}{Section}{Sections}

\usepackage{grffile}

\newcommand{\Msun}{{\rm M}_{\odot}}

\DeclareSIUnit{\pixel}{pixel}
\DeclareSIUnit{\pixels}{pixels}
\DeclareSIUnit{\degg}{deg}
\DeclareSIUnit[number-unit-product=]{\ahour}{\textsuperscript{h}}
\DeclareSIUnit{\arcsec}{arcsec}
\DeclareSIUnit{\yr}{yr}
\DeclareSIUnit{\pc}{pc}
\DeclareSIUnit{\parsec}{parsec}

\author[A. Guzmán-Ortega et al.]
{
	\parbox{18cm}{
    Alejandro Guzmán-Ortega,$^{1}$\thanks{E-mail: a.guzman@irya.unam.mx} 
    Vicente Rodriguez-Gomez,$^{1}$
    Gregory F. Snyder, $^{2}$
    Katie Chamberlain $^{3}$ \\
    and Lars Hernquist $^{4}$
  }
	\vspace{0.3cm} \\ 
	$^{1}$ Instituto de Radioastronom\'ia y Astrof\'isica, Universidad Nacional Aut\'onoma de M\'exico, Apdo. Postal 72-3, 58089 Morelia, Mexico \\
	$^{2}$ Space Telescope Science Institute, 
  3700 San Martin Drive, Baltimore, MD 21218, USA \\
	$^{3}$ University of Arizona, 933 N. Cherry Ave, Tucson, AZ 85721, USA \\
	$^{4}$ Harvard-Smithsonian Center for Astrophysics, 60 Garden Street, Cambridge, MA 02138, USA 
  }

\title[Morphological signatures of mergers in TNG50 and KiDS]{%
Morphological signatures of mergers in the TNG50 simulation and the Kilo-Degree Survey: the merger fraction from dwarfs to Milky Way-like galaxies}

\begin{document}

\maketitle
\begin{abstract}
Using the TNG50 cosmological simulation and observations from 
the Kilo-Degree Survey (KiDS), we investigate the 
connection between galaxy mergers and optical morphology in the 
local Universe over a wide range of galaxy stellar masses 
($8.5\leqslant\log(M_\ast/\text{M}_\odot)\leqslant11$). To this end,
we have generated over 16~000 synthetic images of TNG50 galaxies 
designed to match KiDS observations, including the effects of dust 
attenuation and scattering, and used the \textsf{statmorph} code 
to measure various image-based morphological diagnostics in the $r$-band for 
both data sets. Such measurements include the Gini--$M_{20}$ 
and concentration--asymmetry--smoothness statistics. 
Overall, we find good agreement between the 
optical morphologies of TNG50 and KiDS galaxies, although the former 
are slightly more concentrated and asymmetric than 
their observational counterparts. Afterwards, we trained a 
random forest classifier to identify merging galaxies in 
the simulation (including major and minor mergers) using the 
morphological diagnostics as the model features, 
along with merger statistics from the merger trees as the ground 
truth. We find that the asymmetry statistic exhibits the 
highest feature importance of all the morphological parameters 
considered. Thus, the performance of our algorithm 
is comparable to that of the more traditional method of selecting 
highly asymmetric galaxies.
Finally, using our trained model, we estimate the 
galaxy merger fraction in both our synthetic and 
observational galaxy samples, finding in 
both cases that the galaxy merger fraction 
increases steadily as a function of stellar mass.
\end{abstract}


\section{Introduction}%
  \label{sec:Introduction}
It is now accepted, within the $\Lambda$ Cold Dark Matter
($\Lambda$CDM) model, 
that galaxy mergers play a fundamental role in
the formation and evolution of galaxies \citep{White1978}. 
These events change the
nature of galaxies in a number of ways: post-mergers exhibit a
diminished
central metallicity \citep[e.g.][]{Kewley2006,Scudder2012};
while merging galaxies not only
have both enhanced morphological disturbances
\citep[e.g.][]{Conselice2003,Lotz2008,Casteels2014} 
and star formation rates \citep[e.g.][]{Patton2013,Thorp2019}
in comparison to non-mergers, but also tend to exhibit active
galactic nuclei \citep[e.g.][]{Ellison2011,Ellison2019}.

Modern cosmological
simulations, such as those from the
EAGLE \citep{Schaye2015,Crain2015}, 
Illustris 
\citep{Vogelsberger2014,Vogelsberger2014a,Genel2014,Sijacki2015}, 
IllustrisTNG 
\citep{Marinacci2018,Naiman2018,Nelson2018,Pillepich2018,
Springel2018,Nelson2019} and Horizon-AGN \citep{Dubois2014} 
projects, 
have been able not only to corroborate most observational
findings about mergers, but also to provide additional 
insights that can be used to inform and interpret observations. 
For example, 
one of the most interesting and
studied topics is the determination of the galaxy merger rate, namely,
the number of mergers per unit time. A proper determination of
this quantity is important
to fully understand the role 
of interactions
in galactic structure and star formation rates, as well as to test 
hierarchical galaxy formation models.


Theoretically, the merger rate is estimated via
semi-empirical (e.g. \citealt{Stewart2009,Hopkins2010}) 
and semi-analytic 
(e.g. \citealt{Guo2008})
models as well as using hydrodynamic cosmological
simulations (e.g. \citealt{Rodriguez-Gomez2015}). 
In particular, \citet{Rodriguez-Gomez2015} used the Illustris 
simulation to quantify the 
galaxy-galaxy merger rate as a function of stellar mass, merger mass 
ratio, and redshift, finding that it increases steadily with 
stellar mass and redshift, while being in good agreement with 
observational constraints for both 
intermediate-sized and massive galaxies. 

On the observational side, it is important to note that the
merger rate cannot be estimated directly. Instead, the merger
fraction, i.e. the fraction of galaxies observed to be 
undergoing a merger, 
must be computed
first, and then translated into a rate dividing by an appropriate
\textit{observability}
time-scale reflecting the merging detection period. 
The merger fraction is typically computed by using
observations of close pairs or morphologically disturbed galaxies.

Close-pair merger candidates have been identified by several
authors 
(e.g. \citealt{Lin2004,Propris2005,Kartaltepe2007,Besla2018,Duncan2019}) 
as galaxies with a neighbour within a small projected angular
separation and with a small line-of-sight relative 
radial velocity. 
Alternatively, since there is a close connection between 
galactic structure and merging processes, some types of
morphologically disturbed galaxies are also considered to be 
ideal merger candidates.

In particular, non-parametric morphological diagnostics, such as the
concentration--asymmetry--smoothness (CAS, \citealt{Conselice2003}), 
Gini--$M_{20}$ \citep{Lotz2004}, and 
multimode--intensity--deviation 
\citep[MID,][]{Freeman2013} statistics, 
have been successfully used to identify galaxy mergers and to 
study, among other things, the evolution of the observational
  merger rate
\citep[e.g.][]{Lotz2011}, its dependence on galaxy stellar mass 
\citep[e.g.][]{Casteels2014}, as well as to obtain the galaxy merger 
rate and merging time-scales from hydrodynamical simulations, 
allowing direct comparisons with observational estimates 
\citep[e.g.][]{Bignone2016,Whitney2021}.

More recently, machine learning and deep learning 
methods have been adopted, for the
same purpose, as an
alternative to more standard methods. For instance,
\citet{Snyder2019} considered a high-mass galaxy sample from the
original Illustris simulation to
create, based on image-based 
morphological calculations and merger statistics, 
a training data set
that was fed to a random forest classifier. The resulting
model was then used to perform observational and theoretical
estimates of the merger rate as a function of redshift. This method was
also recently used to perform merger classifications on JWST-like simulated images \citep{2022arXiv220811164R}.
Similarly, \citet{2019ApJ...872...76N} combined non-parametric morphological statistics 
with a linear discriminant analysis (LDA) classifier to characterize simulated
mergers and found that the LDA classifier outperformed the individual metrics.

Furthermore, convolutional neural networks have been adopted to 
study galaxy morphology in different contexts. To list a few, they
have been applied to differentiate disk galaxies from bulge-dominated
systems in both observations and cosmological simulations 
\citep{HuertasCompany2015,Huertas-Company2019}, 
to identify
merging galaxies in cosmological simulations 
\citep{Ciprijanovic2020}, 
to determine the effect of merging events on the 
star formation rates of galaxies 
\citep{Pearson2019}, to predict the stage of interaction 
on synthetic galaxy 
images and to assess the degree of realism and post-processing 
needed on these 
synthetic images to perform adequate deep learning estimations 
\citep{Bottrell2019}, and to identify high-mass major merger events 
in observations 
and simulations and subsequently estimate the merger fraction 
\citep{2020A&A...644A..87W}. 

Despite the well-established link between mergers and morphology 
in massive galaxies, the incidence and effects of mergers in 
dwarf galaxies $(M_\ast < 10^{9.5}\,\mathrm{M}_\odot)$ 
are more uncertain. This is an important topic to explore, 
since galaxy mergers can be transformative events and 
dwarfs represent the majority of the galaxy population.
For example, \citet{Casteels2014} studied a local galaxy sample 
at $z<0.2$ and 
$10^{8}\,\mathrm{M}_\odot<M_\ast<10^{11.5}\,\mathrm{M}_\odot$ 
to infer
the mass-dependent galaxy merger fraction and merger rate 
by measuring the asymmetry parameter from galaxy images. 
Their estimated major merger fraction is a decreasing function
of stellar mass, falling from $4\%$ at 
$M_\ast\sim10^{9}\,\mathrm{M}_\odot$ to $2\%$ at
$M_\ast\sim10^{11}\,\mathrm{M}_\odot$.
This finding suggests that 
galaxy interactions might become increasingly important for 
lower-mass galaxies.

Similarly, \citet{Besla2018} computed the frequency of companions 
for a low-redshift dwarf galaxy ($0.013<z<0.0252$; $2\times10^{8}\,\text{M}_\odot<M_\ast<5\times10^{9}\,\text{M}_\odot$) sample from the
Sloan Digital Sky Survey (SDSS), comparing it to a
mock galaxy sample from the original Illustris simulation. 
One of the goals of their study was to estimate the 
major pair fraction as a function of stellar mass, finding that this
quantity increases slowly with stellar mass, but
does not follow the decreasing trend reported by
\citet{Casteels2014}. 
Motivated by such opposing results, 
in this paper we revisit the topic of the
mass-dependent merger fraction, with an emphasis on the 
regime of dwarf galaxies. 

In this work, we use the 
TNG$50$ simulation from the 
IllustrisTNG project to investigate the galaxy merger
fraction at the low-mass end ($8.5\leqslant\log(M_\ast/\text{M}_\odot)\leqslant11$),
and whether it can be inferred  
statistically from morphological disturbances in large 
samples of galaxies. 
For this purpose, we generate a large set of 
synthetic images of TNG$50$ galaxies, including the effects of dust 
attenuation and scattering,  
designed to match observations 
from the Kilo-Degree Survey (KiDS). 
We then calculate several image-based morphological 
statistics for both the real and simulated galaxy samples, 
and explore the connection between morphology and mergers in 
the simulation. 
Instead of solely using the asymmetry statistic as a merger 
indicator, 
we follow a similar approach to that of \citet{Snyder2019} 
and train a random forest classifier 
using several non-parametric morphological indicators 
as the model features and merger statistics 
from the simulation merger trees as the ground truth. 
The trained models are then applied to the observational 
sample in order to estimate the local merger fraction as a 
function of galaxy 
mass in the real Universe.

This paper is structured as follows. In \cref{sub:The IllustrisTNG
simulations} we briefly review
the IllustrisTNG simulations, and in \cref{sub:Observational
sample,sub:Simulated sample} we describe the observational and
simulated galaxy samples used in
our analysis. 
In \cref{sub:Synthetic image generation,sub:Source segmentation} 
we describe the design and generation of synthetic
images from the simulated sample, as well as the source 
segmentation and deblending procedures. 
Subsequently, in \cref{sub:Morphological_measurements} 
we present the morphological diagnostics measured
on both galaxy samples. In 
\cref{sub:Merger identification} we define our merging and 
non-merging 
simulated samples, and in \cref{sub:Random
forest classification} we describe the training and calibration 
of our random forest classifier.
Our main results are given in \cref{sec:results}, where we 
examine the morphological differences between our observational 
and synthetic galaxy samples 
(\cref{sub:The morphologies of TNG50 galaxies}) as well as 
between merging and non-merging 
simulated galaxies 
(\cref{sub:The morphologies of intrinsic mergers}), evaluate 
the performance of our random forest classifier 
(\cref{sub:Random forest classification performance}), 
and present the resulting mass-dependent merger fraction 
(\cref{sub:The merger incidence of GAMA-KiDS observations}).
Finally, we discuss our results in \cref{sec:Discussion} and present our conclusions 
in \cref{sec:Summary}.

\section{Methodology}%
  \label{sec:Methodology}
\subsection{The IllustrisTNG simulations}%
  \label{sub:The IllustrisTNG simulations}
The IllustrisTNG project is a suite of $N$-body
magneto-hydrodynamical cosmological simulations that model dark and
baryonic matter assuming a $\Lambda$CDM framework
\citep{Marinacci2018,Naiman2018,Nelson2018,Pillepich2018,Springel2018,Nelson2019}.
The simulation suite consists of three cubic volumes with 
periodic boundary conditions: TNG50, TNG100, and TNG300, 
which measure 51.7, 110.7, and 302.6 Mpc on a side, respectively.

In this work we use the highest resolution version of the TNG$50$
simulation \citep{2019MNRAS.490.3196P,2019MNRAS.490.3234N}, which has a volume of
$\qtty(\SI{51.7}{\mega\pc})^3$ at a baryonic (dark) mass resolution
of $8.5\cdot10^4\,\mathrm{M}_\odot$ ($4.5\cdot10^5\,\mathrm{M}_\odot$)
and a spatial resolution (effectively set by the gravitational 
softening length of stellar and DM particles) of 
${\sim}\SI{300}{\pc}$ at $z=0$.
The starting redshift of the simulation is $z=127$ which is
evolved 
down to $z=0$. The assumed cosmological parameters, obtained from 
\citet{Ade2016}, are $\Omega_{\Lambda,\,0}=0.6911$,
$\Omega_{m,\,0}=0.3089$, $\Omega_{b,\,0}=0.0486$,
$\sigma_8=0.8159$, $n_s=0.9667$ and $h=0.6774$.

The galaxy formation model in IllustrisTNG includes prescriptions 
for gas radiative cooling, star formation and evolution, supernova 
feedback, metal enrichment, and feedback from supermassive black 
holes \citep[see][for a full description]{Weinberger2017a,2018MNRAS.473.4077P}.
This model was tuned to approximately 
match several 
observational properties, such as the star formation rate density
at $z=0-8$, the galaxy mass function and sizes at $z=0$, the
stellar-to-halo and BH-to-halo mass relations at $z=0$ and the gas
mass fraction within galaxy clusters \citep{2018MNRAS.473.4077P}. 
Since the simulations were not adjusted to match
galaxy morphology, it is noteworthy that there is a reasonable level
of morphological consistency
between TNG$100$ galaxies and a comparable observational
sample \citep{Rodriguez-Gomez2019}, as well as other properties of galaxies such as
their star formation activity \citep{2019MNRAS.485.4817D}, 
resolved star formation \citep{10.1093/mnras/stab2131}, and metallicities
\citep{2019MNRAS.484.5587T}. 

In order to identify DM haloes in the simulation, 
friends-of-friends groups are constructed using
the percolation algorithm by \citet{Davis1985}, linking dark
matter particles based on their inter-particle separation. The
linking length used in the simulations is $b=0.2$ (in units of 
the mean interparticle distance). Furthermore,
subhaloes are identified with the \textsc{\textsf{subfind}} algorithm
\citep{Springel2001a,Dolag2009}.


\begin{figure}
  \begin{center}
    \includegraphics[scale=0.6]
    {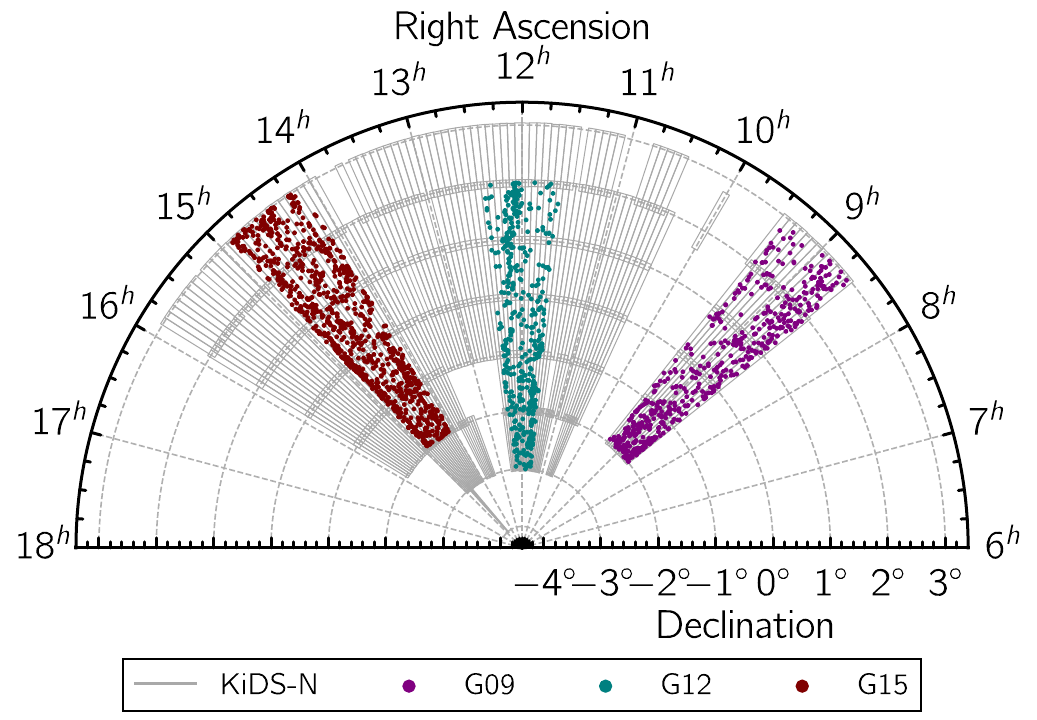}
  \end{center}
  \caption[Position of GAMA galaxies with respect to KiDS-N tiles.]
  {Position of GAMA galaxies with respect to KiDS-N tiles. 
   This figure demonstrates that the GAMA sample is fully
   contained within KiDS. Galaxies shown here constitute our final
   GAMA-KiDS sample with $z<0.05$ and 
   $8.5\leqslant\log \qtty( M_\ast/\mathrm{M}_\odot)\leqslant11$.}
   \label{fig:gamakidstiles}
\end{figure}
\begin{figure}
  \begin{center}
    \includegraphics[scale=0.45]{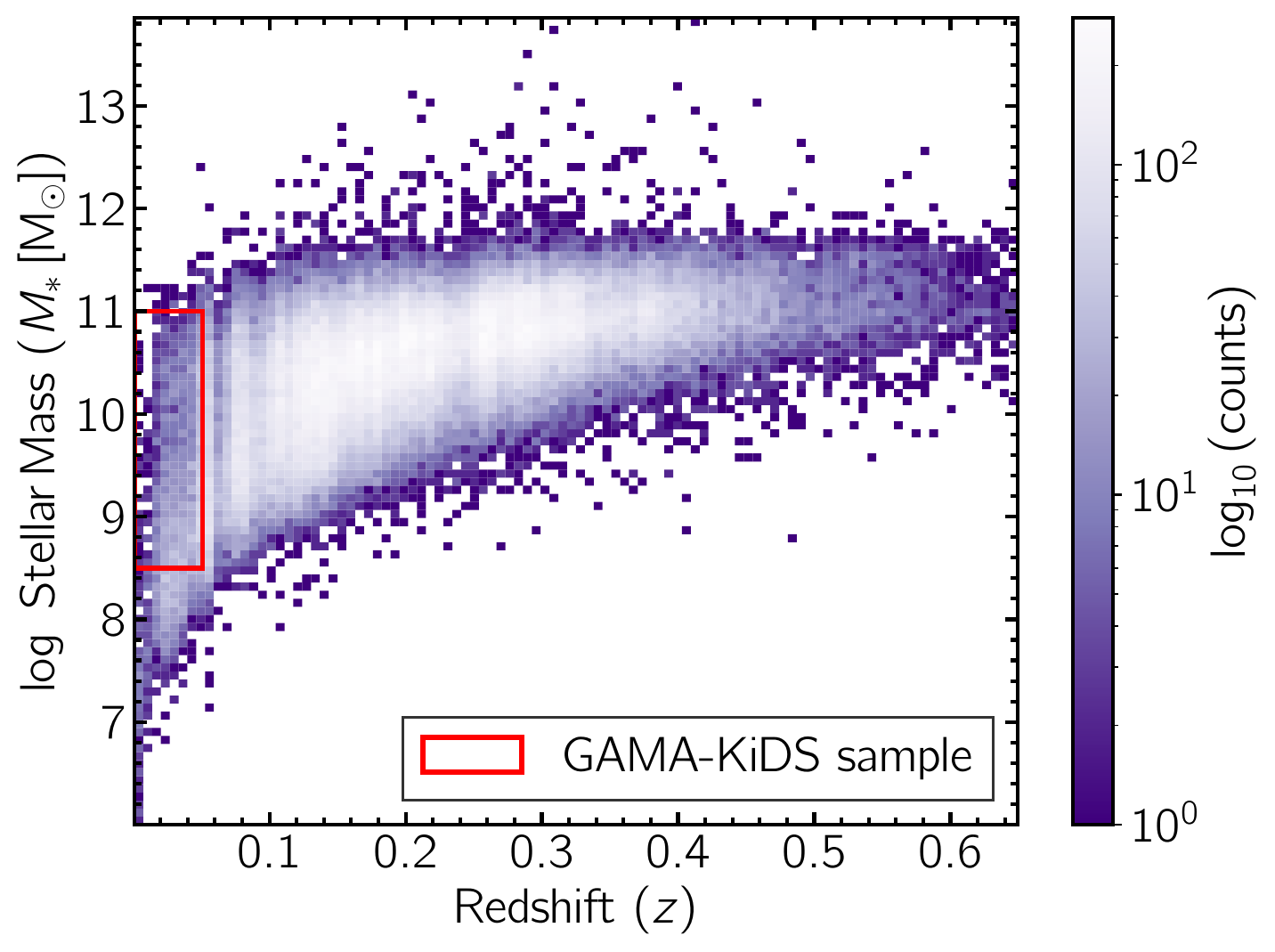}
  \end{center}
  \caption[Distribution of stellar masses and redshifts of GAMA 
  galaxies.]{Distribution of stellar masses and redshifts of GAMA
             galaxies. The rectangle encloses the examined 
             ($z<0.05;\, 8.5\leqslant\log \qtty( M_\ast / 
             \mathrm{M}_\odot)\leqslant11$) GAMA-KiDS sample.}
  \label{fig:z_vs_mass}
  \end{figure}

\subsection{Observational sample}%
  \label{sub:Observational sample}
The Kilo-Degree Survey (KiDS; \citealt{deJong2013}) 
is an ongoing optical wide-field imaging survey operating 
with the OmegaCAM camera at the Very Large Telescope 
(VLT) Survey Telescope, whose main goal is to map 
the Universe's large-scale matter distribution using weak 
lensing shear and photometric redshift measurements. 

KiDS is divided into two patches, one in the 
north (KiDS-N) and the other in the south (KiDS-S). 
The first of these is found near the  equator in the 
Northern Galactic Cap, while the second is found
around the South Galactic Pole; in combination, they cover 
$\SI{\sim1350}{\degg\squared}$ of the sky. 
The analysis presented in this work was conducted using products
from the fourth data release of KiDS \citep{Kuijken2019}. 
The survey has a typical seeing of 
\ang[angle-symbol-over-decimal=true]{;;0.7} with
$5\sigma$
depths of 24.2, 25.1, 25.0 and 23.7 mag
for the filters \emph{u}, \emph{g}, \emph{r} and \emph{i},
respectively. We only considered $r$-band 
stacked
images (pixel scale of \SI{0.2}{\arcsec\per\pixel}) from KiDS-N.

Our galaxy sample was constructed using data from the 
Galaxy and Mass Assembly (GAMA; \citealt{Baldry2018})
survey, a large catalogue of galaxies with reliable redshifts 
obtained from spectroscopic and multi-wavelength observations.
The Anglo-Australian Telescope's (AAT) AAOmega multi-object 
spectrograph
was used to conduct the survey, which covered three equatorial 
(G$09$, G$12$ and G$15$) and two southern regions (G$02$ and G$23$).
The first three, chosen for this study, each
encompassed \ang{\sim5} in declination and \ang{\sim12} in right
ascension, and were centred at roughly \SI{9}{\ahour} (G$09$),
\SI{12}{\ahour} (G$12$) and \SI{15}{\ahour} (G$15$). 
\cref{fig:gamakidstiles} 
shows the locations of galaxies from regions G$09$, 
G$12$ and G$15$ in relation to tiles from KiDS-N,
revealing that
practically all of these galaxies are included within the 
KiDS footprint. 
This makes it straightforward to match GAMA galaxies to KiDS 
tiles and create cutouts centred on any individual galaxy. Our 
 \begin{figure*}
  \begin{center}
    \includegraphics[width=\textwidth]{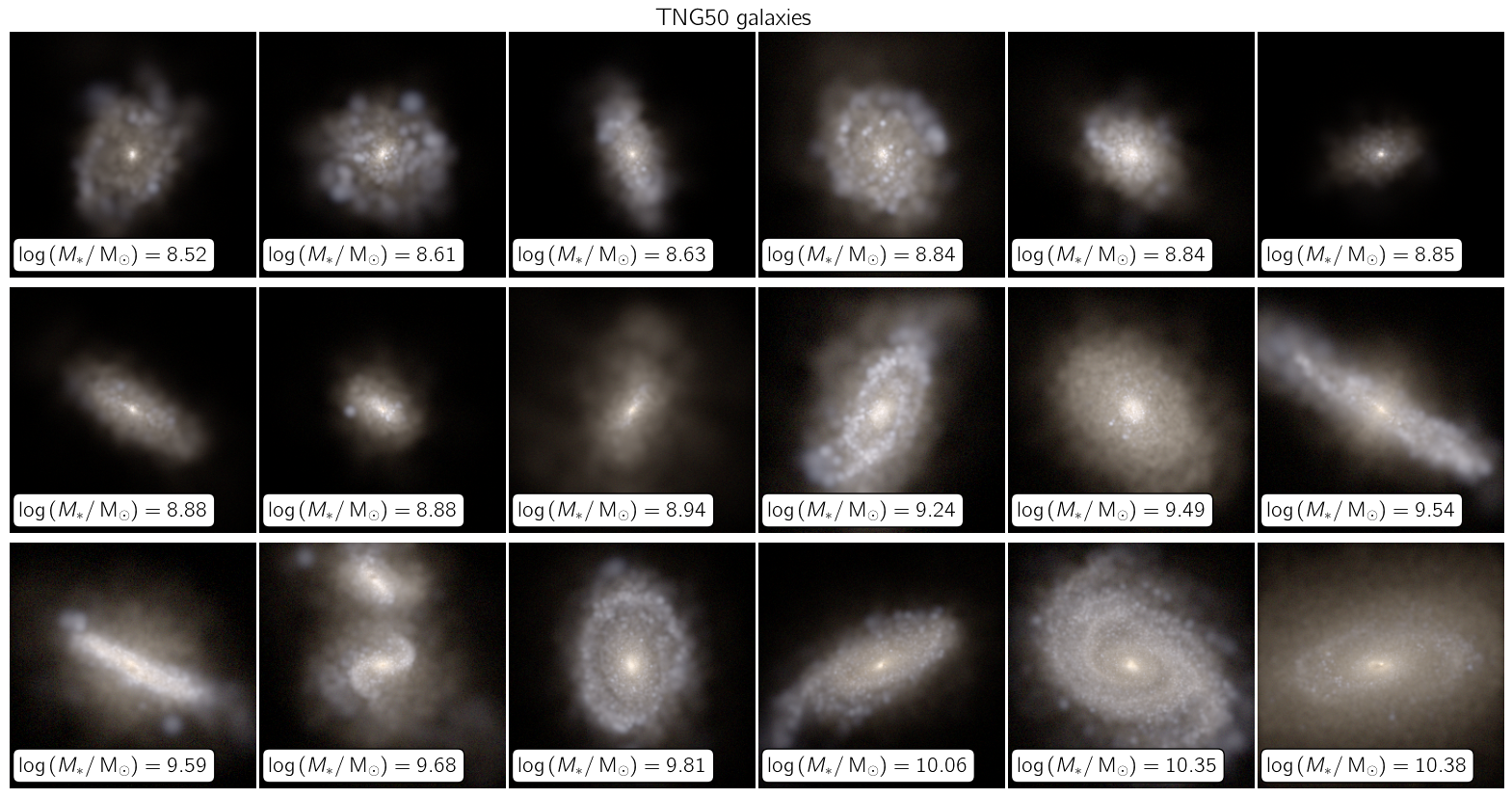}
  \end{center}
  \caption{Composite $g,r,i$ idealized synthetic images from the simulated TNG50 sample, as obtained with the process described in \cref{sub:Synthetic image generation}. Realistic $r$-band counterparts are shown in the first three rows of \cref{fig:sim_vs_obs_synth_images}. Labels denote the stellar mass of each galaxy.}
  \label{fig:comp_images}
  \end{figure*}
\begin{figure*}
  \begin{center}
    \includegraphics[width=\textwidth]{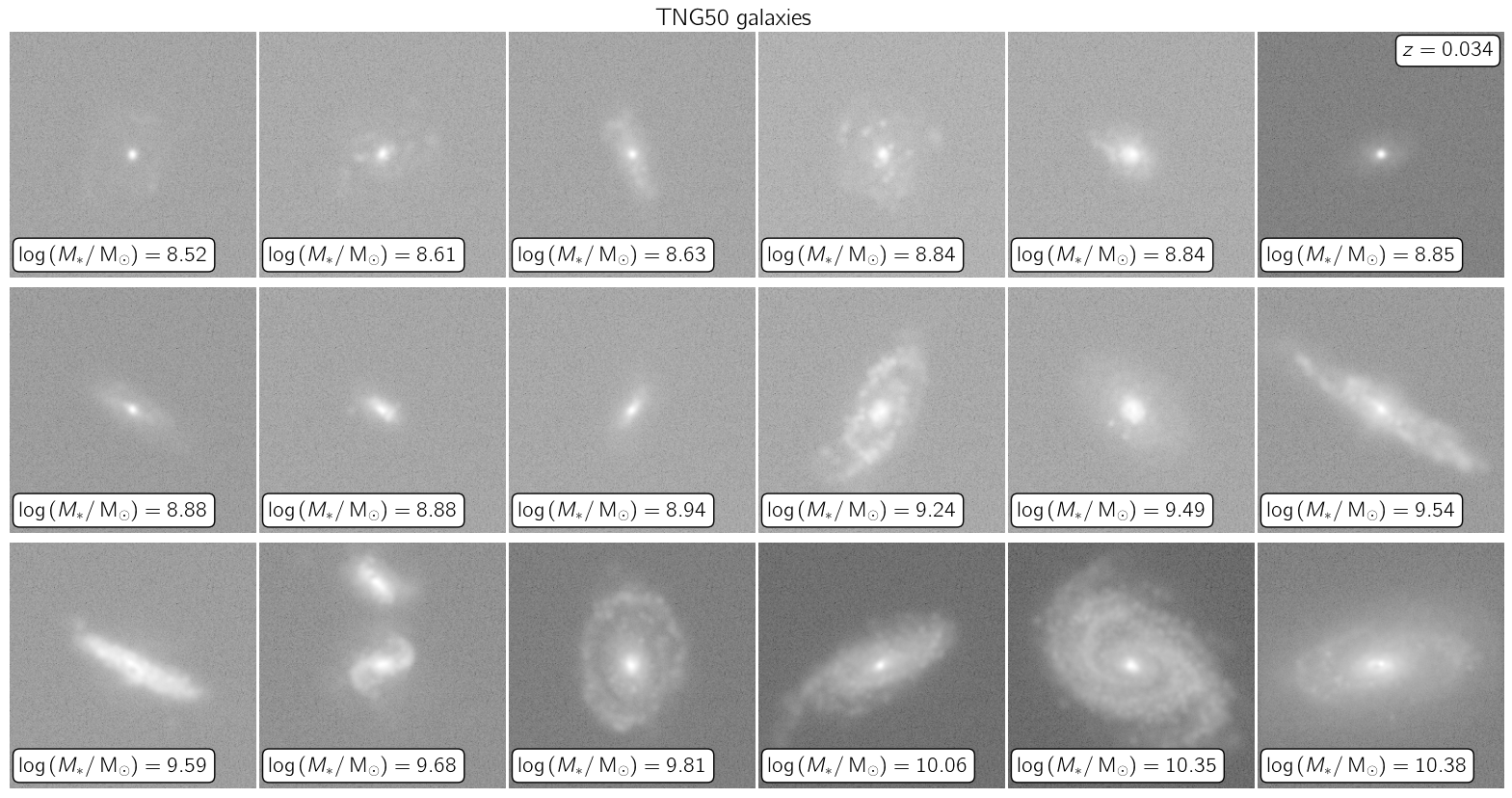}\\
    \includegraphics[width=\textwidth]{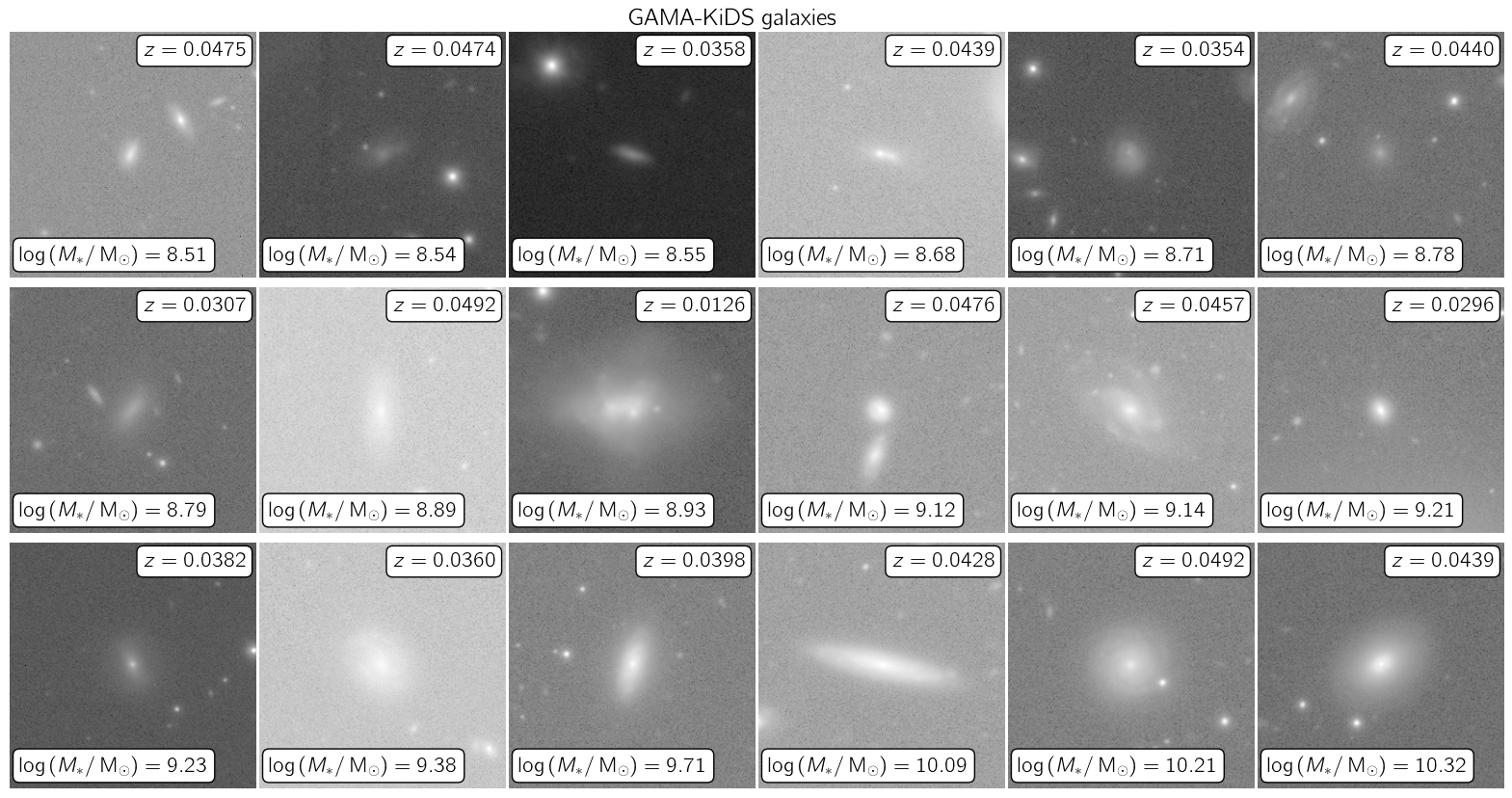}
  \end{center}
  \caption{First three rows: $r$-band synthetic images of TNG50 
    galaxies after applying realism (convolution with a PSF 
    and addition of shot and background noise) to the 
    underlying idealized images, as described in  \cref{sub:Synthetic image generation}, 
    with the labels indicating 
    the corresponding stellar masses. 
    Last three rows: GAMA-KiDS $r$-band galaxy images
  with upper (lower) labels indicating their redshift (stellar 
mass).}
  \label{fig:sim_vs_obs_synth_images}
  \end{figure*}

\noindent position  
matching approach yields a sample of $104~993$ objects with KiDS 
imaging
from an initial sample of $105~474$ GAMA galaxies.

From GAMA's third data release
\citep{Baldry2018} we use the \textsf{StellarMasses} 
\citep{Taylor2011} and
\textsf{SpecCat} \citep{Liske2015} 
products to obtain the stellar mass, redshift and location 
(right ascension and declination) for the
examined galaxies. The distribution of stellar masses
and redshifts of GAMA galaxies can be seen in \cref{fig:z_vs_mass}. 
For this study, we selected galaxies in the mass range 
$8.5\leqslant\log \qtty( M_\ast/\mathrm{M}_\odot)\leqslant11$
with redshift $z<0.05$ in order to
perform a direct comparison to a single snapshot from the TNG$50$
simulation. Our final
GAMA-KiDS sample consisted of $1238$ objects,
and is illustrated in both \cref{fig:gamakidstiles,fig:z_vs_mass}.

Finally, having defined our galaxy sample based on the GAMA 
catalogues, KiDS cutouts for individual galaxies were created 
using utilities
from the \textsf{astropy} library 
\citep{Robitaille2013,Price-Whelan2018}. Specifically, the function
\texttt{match\_coordinates\_sky} was used to locate the nearest KiDS
tile that contained a given GAMA galaxy. Then, the function 
\texttt{Cutout2D} was employed to create individual images with a
fixed size of $240\times240$ pixels. Overall, $104~993$ cutouts were 
obtained from this matching procedure.

\subsection{Simulated sample}%
  \label{sub:Simulated sample}
We consider galaxies from the TNG50 simulation with
stellar masses in the range $8.5\leqslant\log(M_\ast/\text{M}_\odot)\leqslant11$, 
which is consistent with our GAMA-KiDS sample, from a single simulation 
snapshot at 
$z=0.034$ (snapshot $96$) that is close to the median 
redshift of the galaxies in our observational sample. The 
resulting mock galaxy catalogue consists of $5561$ galaxies.
\begin{figure*}
  \begin{center}
    \includegraphics[width=\textwidth]{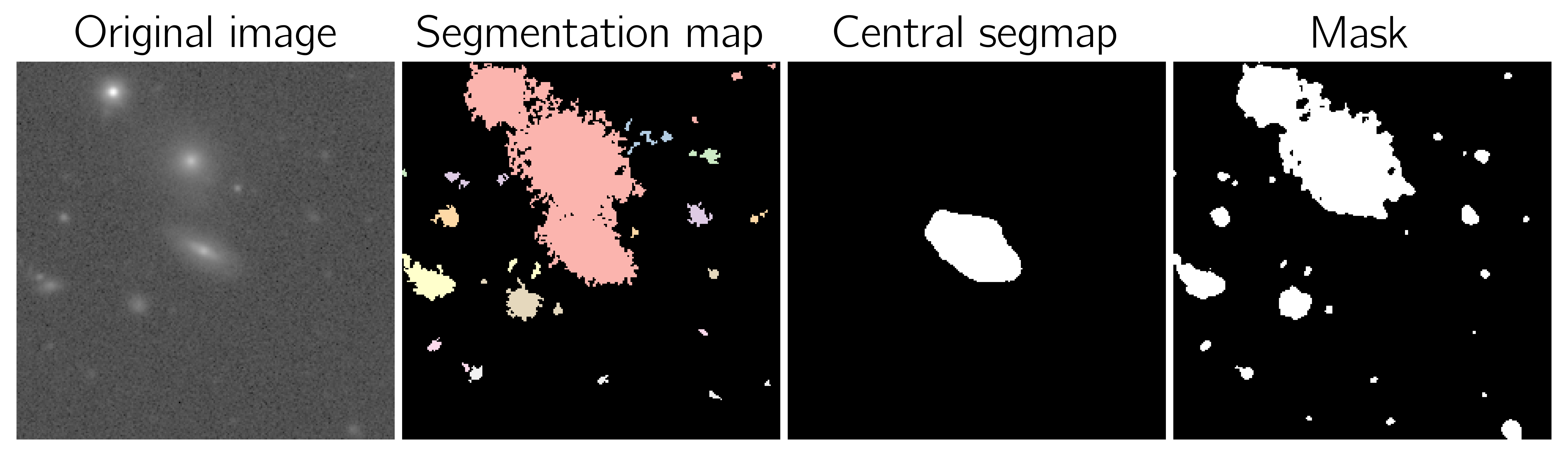}
  \end{center}
  \caption{Segmentation procedure. Leftmost panel: real KiDS $r$-band 
  image showing multiple sources; second from left: segmentation map
  obtained from the previous image; 
  second from right: regularised segmentation map for the object 
  of interest (central galaxy); rightmost entry: regularised mask.}
  \label{fig:deblending}
  \end{figure*}

\subsection{Synthetic image generation}%
  \label{sub:Synthetic image generation}
Galaxy images were constructed from the light distributions of
stellar populations, including 
dust effects such as scattering and attenuation.
Following \citet{Rodriguez-Gomez2019}, we use   
different image generation pipelines 
based on the
value of the star-forming gas fraction ($f_{\mathrm{ gas,\,sf }}$) 
in simulated galaxies:
synthetic images for galaxies with $f_{\mathrm{ gas,\,sf }} < 0.01$ 
were generated using the \textsc{\textsf{galaxev}}
stellar population synthesis code \citep{Bruzual2003}, 
while galaxies with $f_{\mathrm{ gas,\,sf }} \geqslant 0.01$
additionally model young stellar populations (age $<$ 10 Myr) with the
\textsc{\textsf{mappings-iii}}  libraries \citep{Groves2008} 
and include dust radiative transfer using the 
\textsc{\textsf{skirt}} code \citep{Baes2011,Camps2015}. 
We note that we avoid processing low-gas galaxies with 
\textsc{\textsf{skirt}} simply for performance reasons, 
and that both pipelines would produce essentially indistinguishable
images for such objects (further details can be consulted in
\citealt{Rodriguez-Gomez2019}).

For both pipelines, the light contribution from each stellar 
particle was smoothed using a particle hydrodynamics spline kernel 
\citep{Hernquist1989,Springel2001} with an 
adaptive smoothing scale given, for each
simulation particle, as the three-dimensional distance to 
the $32$nd nearest neighbour. Furthermore, the synthetic images
were created with the pixel scale of KiDS 
(\SI{0.2}{\arcsec\per\pixel}) and were mock-observed 
from the Cartesian projections \emph{xy, yz} and
\emph{zx}, also setting the field of view of each object to
$240\times240$ pixels, and taking cosmological effects into 
account (such as surface brightness dimming) 
assuming they are located at $z = 0.034$. This procedure yielded idealised 
images in four broadband filters corresponding to the $g,r,i,z$ bands, of 
which we exclusively used the $r$-band for our analysis. We point out that galaxies
from the $xy$ projection were used in the discussion presented in 
\cref{sub:The morphologies of TNG50 galaxies,sub:The morphologies of intrinsic mergers}, while
galaxies from all three projections were employed from \cref{sub:Random forest classification performance}
onwards. The units of the synthetic images are analog-to-digital units (ADU) 
per second, consistent with real KiDS science images. 
\cref{fig:comp_images} shows 
composite images of randomly selected galaxies from the simulated sample, presented in order of increasing stellar mass, using the KiDS $g,r,i$ filters.

Finally, realism was added to the idealised images via 
convolution with a
point spread function (PSF) and addition of shot and uniform background noise. 
We convolved each image with a 2D Gaussian 
PSF with full width at
half maximum (FWHM) 
equal to \ang[angle-symbol-over-decimal=true]{;;0.7},
which corresponds to the median PSF from KiDS $r$-band images. Shot noise 
was included by assuming an effective gain of $3\times10^{13}$ 
electrons per data unit, also consistent with KiDS data products, 
while background noise was modelled 
as a Gaussian random variable 
with uniform standard deviation 
$\sigma_{\mathrm{ bkg }}=2\times10^{-12}\,\mathrm{ADU\,s^{-1}}$ 
across each simulated image. The first three rows from \cref{fig:sim_vs_obs_synth_images} show the same synthetic images of \cref{fig:comp_images} after
adding realism, which are
compared in the next three rows to randomly selected galaxies from observations.
   \begin{table}
\caption{Galaxy deblending and segmentation parameters used by 
           \textsc{\textsf{sep}}.}
  \centering
  \bgroup
  \def\arraystretch{1.2}
  \begin{tabular}{ccc}
  \toprule
  \text{Parameter} & \text{Description} & \text{Value} \\ 
  \midrule 
  \texttt{\textbf{minarea}} & \text{Minimum galaxy area} & 10 pixels
   \\
  \texttt{\textbf{deblend\_nthresh}}& \text{Number of deblending levels} &
   16 \\
  \texttt{\textbf{deblend\_cont}} & \text{Minimum deblending contrast} &
   0.0001 \\
  \texttt{\textbf{thresh}} & \text{Minimum detection threshold} & 0.75 \\
  \bottomrule
  \end{tabular}
  \egroup
  \label{tab:sepparams}
  \end{table}

\subsection{Source segmentation and deblending}%
  \label{sub:Source segmentation}
Segmentation is the process in which distinct sources within an
astronomical image are labelled with different integer values,
reserving the zero-value for the background. 
The resulting array, with
the same shape as the original image, is called a segmentation map 
or segmentation image.

Proper segmentation is a crucial step for the morphological
measurements presented in \cref{sub:Morphological_measurements}, 
since it defines the region that corresponds to the 
galaxy of interest while removing contaminant objects. The
most challenging stage of this procedure is deblending, i.e.
the separation of two or more overlapping different sources. 

Since we are interested in morphology-based merger identifications 
on individual galaxies (instead of, for example, counting
close pairs)
and because it is almost impossible, based on
imaging alone, to distinguish between true companions and
contaminants (such as chance projections along the line of sight),
deblending was applied to all examined sources.

Thus, we have created segmentation maps for both the observational 
and mock
samples using \textsc{\textsf{sep}} \citep{Barbary2016}, 
a \textsf{Python} library that implements the core functionality 
of \textsf{SE}\textsc{\textsf{xtractor}} \citep{Bertin1996}. 
We have controlled source detection and deblending using the following
\textsc{\textsf{sep}} input parameters: \texttt{\textbf{thresh}}, 
the detection threshold in standard deviations;
\texttt{\textbf{minarea}}, the minimum number of pixels in an
object; \texttt{\textbf{deblend\_nthresh}}, the number of deblending
levels, and \texttt{\textbf{deblend\_cont}}, 
the minimum contrast ratio
for source deblending. The values of these input parameters are given
in \cref{tab:sepparams}.

Based on segmentation maps, we produced a mask for each galaxy.
This was accomplished by identifying the object that coincided with
the galaxy of interest and labelling it as the main source; the remainder of
the segments, excluding the background, constituted the mask. It is 
worth mentioning that the main segment and the mask were 
\emph{regularised} in the sense that they were smoothed by a uniform 
filter with a size of $10\times10$ pixels. 
\cref{fig:deblending} shows a schematic of this procedure.

\subsection{Morphological measurements}%
  \label{sub:Morphological_measurements}
Morphology calculations were done using \textsf{statmorph} 
\citep{Rodriguez-Gomez2019}, a \textsf{Python} package for 
computing non-parametric morphological diagnostics of galaxy images,
as well as fitting 2D Sérsic profiles. 
In order to run the code we used the science
images,
their
segmentation maps and their associated masks, as well as the
\texttt{\textbf{gain}} 
factor, a scalar that converts the image units into 
$e^{-}\,\text{pixel}^{-1}$, using the same value as in 
\cref{sub:Synthetic image generation}.

In this work we consider the following parameters: concentration,
asymmetry and smoothness (CAS;
\citealp{,Conselice2003}), Gini-$M_{20}$ statistics
\citep{Lotz2004,Snyder2015a,Snyder2015b}, 
multimode, intensity and deviation 
(MID;
\citealp{Freeman2013}) and variations of the asymmetry parameter,
such as the outer ($A_O$; \citealp{Wen2014}) and shape ($A_S$;
\citealp{Pawlik2016}) asymmetries.  
Below we briefly describe each of them.

The concentration parameter ($C$; 
\citealt{Bershady2000,Conselice2003}) 
is a measure of the quantity of 
light at a galaxy's centre in comparison to its outskirts and is given
by $5 \log(r_{80}/r_{20})$, where $r_{20}$ and $r_{80}$ are the
radii of circular apertures containing 20\%
($r_{20}$) and
80\% ($r_{80}$)
of the galaxy's total flux. 
Elliptical galaxies exhibit high concentration ($\sim4$) values, whereas spiral galaxies 
have smaller ($\sim3$) values.

The asymmetry index ($A$; 
\citealt{Abraham1996x,Conselice2000,Conselice2003}) 
is calculated by subtracting a
galaxy image from its $ { 180 }^{ \circ } $-rotated
counterpart, and
is used as a measure of what fraction of light is due to 
non-symmetric components.
The equation for computing this parameter is given by
  \begin{equation}
   A = \frac{ \sum_{i,\,j}\abs{I_{ij} - I_{ij}^{180}} }{ \sum_{i,\,j} \abs{I_{ij}}} - A_{ \textsc{bkg} }, \label{eq:asymmetry} 
  \end{equation} 
where $ I_{ij} $ and $ { I }_{ij}^{ 180 } $ are, respectively, 
the pixel flux values of the original and rotated distributions,
and $ A_ \textsc{bkg} $ is the average asymmetry of the
background. High asymmetry values are often used 
to identify possible recent interactions and galaxy mergers.
\begin{table*}
  \centering
  \bgroup
  \def\arraystretch{1.1}
  \begin{tabular}{ccccccccccc} 
  \toprule
  \multirow{2}{*}{Combination} & \multicolumn{10}{c}{Parameter} \\
  & $C$ & $ A $ & $ S $ & $F\qtty(G,\,M_{20})$ & 
  $S\qtty(G,\,M_{20})$ & $M$ &  $I$ &  $D$ &  $A_O$ &  $A_S$ \\
  \midrule 
  1 &\textbf{Yes} & \textbf{Yes} &\textbf{Yes} 
    &\textbf{Yes}&\textbf{Yes} &No&No&No&No&No \\
  2 &\textbf{Yes} & \textbf{Yes} &\textbf{Yes} 
    &\textbf{Yes}&\textbf{Yes} &\textbf{Yes} &\textbf{Yes}
    &\textbf{Yes} &No&No \\
  3 &No & \textbf{Yes} &\textbf{Yes} 
    &No&No &No &No
    &No &\textbf{Yes} &\textbf{Yes}  \\
  \bottomrule
  \end{tabular}
  \egroup
  \caption[Feature combinations considered during
  training.]{Feature combinations considered during hyper-parameter
  tuning and cross-validation.}
  \label{tab:comb_features}
  \end{table*}

The smoothness parameter ($S$; \citealt{Conselice2003}) is estimated
in a similar way to the
asymmetry, by 
subtracting a
galaxy distribution from a counterpart that has been smoothed 
by a boxcar filter of width $ \sigma $, and it indicates the 
fraction of light that is contained in 
\emph{clumpy} regions (e.g. high frequency disturbances). Following 
\citet{Lotz2004}, we set the value of $\sigma$ to $25\%$ of the 
Petrosian radius.

The Gini index ($G$; \citealt{Abraham2003,Lotz2004}) 
quantifies the degree of inequality of 
the brightness distribution in a set of pixels. 
The Gini coefficient
is equal to $1$ when all the galaxy light is concentrated in one
pixel; conversely, it is equal to zero when the light distribution is
homogeneous across all pixels.
Early-type galaxies, as well as galaxies with one or more bright nuclei, exhibit high Gini values.

The $ M_{\mathrm{ 20 }} $ coefficient is the normalised second 
moment of a galaxy’s brightest regions, containing $20\%$ of 
the total flux. Mergers and star-forming disk galaxies tend to have 
high $M_{20}$ values.

The bulge parameter ($F(G,\,M_{20})$; \citealt{Snyder2015b}) is a 
linear combination of $G$ and 
$M_{\mathrm{ 20 }}$. In the $G$-$M_{\mathrm{ 20 }}$ space,
early-type, late-type and merging galaxies are found by the 
position they occupy relative to two intersecting lines 
\citep{Lotz2008}. The bulge
parameter, $F(G,\,M_{20})$, is then defined as the position
along the line with origin at the intersection
$ \qtty( G_0=0.565,M_{20,\,0}=-1.679 ) $ 
that is perpendicular to the line that separates early-type and
late-type galaxies, scaled by a factor of $5$,
  \begin{equation}
   F \qtty( G,\,M_{20} ) = -0.693M_{20}+4.95G-3.96. \label{eq:} 
  \end{equation}
The merger parameter ($S(G,\,M_{20})$, \citealt{Snyder2015a})
has a similar definition to $F \qtty( G,\,M_{20} )$. It is given as
the 
position along a line with origin at  $ \qtty( G_0,M_{20,\,0} ) $ that
is perpendicular to the line that separates mergers from non-mergers,
  \begin{equation}
   S \qtty( G,\,M_{20} ) = 0.139M_{20}+0.990G-0.327 \label{eq:}. 
  \end{equation} 

The multimode ($M$; \citealt{Freeman2013}) parameter is the 
pixel ratio of the two brightest
regions of a galaxy, which are identified following a 
threshold method:
bright regions are identified when they are above a threshold value, 
a process repeated for different values until the ratio is 
the largest. Double-nuclei systems tend to have values close to one.

The intensity ($I$; \citealt{Freeman2013}) parameter is the flux
ratio between the two brightest
subregions of a galaxy. For its computation the watershed algorithm is
used, i.e. the galaxy image is divided into groups such that each
subregion consists of all pixels whose maximum gradient paths lead to
the same local maximum. Clumpy systems often exhibit high intensity 
values.

The deviation ($D$; \citealt{Freeman2013}) parameter is given as 
the normalised 
distance between the 
image centroid and the centre of the brightest region found during 
the computation of the $I$-statistic, and is used to 
quantify the offset between bright regions 
of a galaxy and the centroid.

The outer asymmetry ($A_O$; \citealt{Wen2014}) parameter is
defined in the same way as the conventional asymmetry 
(see \cref{eq:asymmetry}), with the
exception that pixels from the inner elliptical aperture that contains
$50\%$ of the galaxy’s light are not included in the computation.
Pixels outside this area build up the outer half-flux region, for
which
$A_O$ is estimated.
Lastly, the shape asymmetry ($A_S$; \citealt{Pawlik2016}) parameter
is also calculated in the same way as the standard asymmetry, with 
the difference that the measurement is done over a binary
segmentation map 
rather than the galaxy brightness distribution.

Finally, \texttt{\textbf{statmorph}} provides a
\texttt{\textbf{flag}} quality parameter to distinguish between 
reliable and unsuccessful measurements, which are labelled with 
\texttt{\textbf{flag} == 0} and \texttt{\textbf{flag} == 1},
respectively. We find that the overall fraction of flagged galaxies
is low, representing ${\lesssim}10\%$ for both our observational and
simulated samples.
From this point onwards
we only consider galaxies with reliable morphological measurements,
also imposing, for each object, a
minimum signal-to-noise ratio $\expval{S/N}\geqslant2.5$.

\subsection{Merger identification}%
  \label{sub:Merger identification}
Galaxy mergers in the TNG50 simulation can be identified
using merger trees created using the \textsc{\textsf{sublink}} code
\citep{Rodriguez-Gomez2015}. The idea behind the merger trees 
is to associate a given subhalo with its
progenitors and descendants from adjacent snapshots, 
in such a way that a merger event occurs
when a subhalo has two or more different progenitors.
From the merging history catalogues of the TNG50 simulation, 
we have determined which galaxies from our synthetic sample have
experienced a merger within a given period. 

The merger mass ratio is defined as $\mu = M_2 / M_1$, where $M_1$ 
and $M_2$ are the stellar masses of the primary and secondary
progenitors, respectively, measured at the moment when the secondary
progenitor reaches its maximum stellar mass 
\citep{Rodriguez-Gomez2015}. 
Traditionally, major and minor mergers are defined as those with $\mu > 1/4$ and $1/10 < \mu < 1/4$, respectively. 
Throughout this paper we consider a combined sample of 
major + minor mergers, 
with mass ratios $\mu > 1/10$, for all our computations. 
The main reason for this choice is to have a larger training 
sample for our classifier, but it also has the advantage 
of potentially detecting merger signatures that are more subtle 
or long-lasting than those produced by major mergers. 
As discussed in \citet{Lotz2011}, 
both the Gini--$M_{20}$ statistics and the asymmetry parameter are
sensitive to minor mergers as well as major ones. 

In this context, our intrinsic merger sample is
composed of mergers that occurred within ${\pm}0.5$ Gyr 
relative to the reference
redshift ($z=0.034$; snapshot 96).
For definiteness, we note that this time 
window includes mergers that were recorded in the snapshots
94 to 99 (since for a merger event recorded at snapshot 
$k$, the merger must have actually taken place at some
time between snapshots $k-1$ and $k$), 
and represents an observability time-scale of $\approx 1$ Gyr.
From a sample of 15~463 galaxy images with successful morphological
measurements, we identified 833 major + minor mergers within the 
specified time window, representing an overall merger fraction of about 5\%. 

\subsection{Random forest classification}%
  \label{sub:Random forest classification}
The random forest (RF) algorithm \citep{Breiman2001} is an ensemble
method based on independent decision trees, each of them learning 
from subsamples of the input data. Predictions on
unseen data are then given as a majority vote among all
uncorrelated models given by the trees in the forest.
The goal of the algorithm is to learn a rule from the
galaxy inputs $\vb{x}$ (morphological measurements) to the labels 
$y$ (merger statistics), and then generalise over novel inputs.

In this study, feature space is defined by the CAS, Gini-$M_{20}$,
MID, $A_O$ and $A_S$ parameters, representing a total of 
10 features, with inputs given as values of these attributes for the galaxies. 
In other words, inputs can be seen as
vectors $\vb{x}_i$ having different morphological values in each
entry. Likewise, the target values of the algorithm are given by
the merger label $y_i$ of each galaxy, with $y_i=1$ if the 
given input
is a true intrinsic merger (as defined in
\cref{sub:Merger identification}), and 
$y_i=0$ otherwise.

The \textsf{scikit-learn} module \citep{Pedregosa2011} was
used to construct
random forests. The library’s internal implementation of the
classifier is based on the algorithm of \citet{Breiman2001}, 
which incorporates bootstrapping of the training set and randomised
feature selection. Utilities from this and the 
\textsf{imbalanced-learn}
library \citep{Lemaitre2017} were also used. The 
receiver operating characteristic (ROC) curve
was considered to assess the performance of each model. 

In 
this sense, positive predictive value 
(PPV, also known as purity or precision) 
is the ratio between the true positives 
(TP; true mergers 
selected by the classifier) and the sum of all 
objects selected, which 
are the false positives (FP; non-mergers 
selected by the classification)
and true positives, that is,
\(
    \text{PPV} = \text{TP}/(\text{TP}+\text{FP}).    
\) Similarly, true positive rate (TPR, also known 
as completeness or recall) is the ratio between 
TP and the total number of
intrinsic mergers, which is the sum of true positives and 
false negatives (FN; true mergers rejected by the classifier), i.e.
\(
    \mathbf{\text{TPR} = \text{TP}/(\text{TP}+\text{FN}).}
\)
The ROC curve is a plot of the 
TPR against 
the false positive rate (FPR) at various threshold values, 
where the latter is computed as 
the ratio of FP and the total number of 
intrinsic non-mergers, 
which in turn are given as the sum of FP and true 
negatives (TN; non-mergers rejected by the algorithm): \( \text{FPR} = \text{FP}/(\text{FP}+\text{TN}) \). The ROC curve of a random classifier (with no predictive power) is a diagonal 
line
from the origin to the point $(1, 1)$ whereas a perfect classifier 
is described by two lines, the first of these starting from the origin to $(0, 1)$ and the second one
from $(0,1)$
to $(1, 1)$, so that it has $\text{TPR} = 1$ and $\text{FPR} = 0$.

Our full learning dataset consists of 15~463 inputs corresponding to
successful morphological
measurements performed on synthetic galaxy images in three 
orientations along the axes of the simulation volume,
each of them having different values, along with their corresponding 
merger labels (ground truth).

We split our full dataset into training and test sets, assigning
$70\%$ of the inputs to the first set,
and the remaining $30\%$ to the second set. 
This division was done in a stratified manner, which
keeps the original proportion of distinct classes 
(merger or non-merger) in both
samples. We point out that in the full
dataset there are ${\sim}18$ non-mergers for each merger. Thus,
class imbalance was offset by randomly 
under-sampling (RUS) the majority class (non-mergers) to bring it to 
the same size as the merger set. This method is implemented in the 
\textsf{imbalanced-learn} library, and was applied before 
fitting the 
random forest. We also tried over-sampling the 
minority class (mergers)
using the Synthetic Minority Over-sampling Technique 
(SMOTE, \citealt{Chawla2002}), 
which resulted in approximately the same 
performance. Therefore, from this point onwards we will only show 
results obtained with the RUS technique.

Similarly, the tuning of hyperparameters (i.e. parameters used to 
control the training process, instead of those derived from it) 
was done by means of
cross-validation on the training set, conducting a 5-fold
stratified scheme. 
This entails randomly dividing the training set
into five folds, retaining
the proportion of mergers and non-mergers, and repeatedly 
training 
the classifier with a
combination of four of these samples while validating (testing) the
remaining one. Cycling
over all splits, the random forest classifier is trained and 
validated on five different subsets of the original training set.
Simultaneously, the hyper-parameters of each forest can be
optimised in a grid search fashion.
In this step the main optimised parameters were the number of trees in
the forest, the
depth of each tree, the maximum number of leaf (terminal) nodes, and
the balancing of the
sample. Accordingly, we used, respectively, the 
\texttt{\textbf{RandomForestClassifier}} function from
{\textsf{scikit-learn}} to tune the \texttt{\textbf{n\_estimators}},
\texttt{\textbf{max\_depth}},
\texttt{\textbf{max\_leaf\_nodes}} and \texttt{\textbf{class\_weight}}
parameters, while keeping the rest of hyper-parameters at 
their default values.

Furthermore, during training we have explored several combinations and 
sub-samples from all morphological attributes 
in order to test how well they 
would perform on their own. \cref{tab:comb_features} 
shows all such combinations
considered in the RF models.
Lastly, the entire procedure was carried out with the
\texttt{\textbf{pipeline}} and \texttt{\textbf{GridSearchCV}}
functions
from the \textsf{scikit-learn} and \textsf{imbalanced-learn} 
libraries, which allow for cross-validation and
exhaustive hyper-parameter tuning
at the same time. The end result of the process 
is a trained model and a
combination of hyper-parameters
that give, according to a particular metric, the 'best' classification 
of the data. Additionally, the classifier provides for each of the 
galaxies considered, a probability score that is used to label them 
(as mergers or non-mergers). As such, this threshold can be varied 
in order to reach a compromise between successful and unsuccessful 
classifications. This is usually done by maximising certain metrics, such as the $F_1$-score or the Matthews correlation coefficient\footnote{The $F_1$-score is defined as the harmonic mean of precision and recall, 
taking values between 0 (worst) and 1 (best).

The Matthews correlation coefficient (MCC) correlates the ground truth with predictions 
in a binary classification and is given by
\[
\text{MCC} = \frac{\text{TP}\times \text{TN}-\text{FP}\times\text{FN}}{\sqrt{(\text{TP}+\text{FP})(\text{TP}+\text{FN})(\text{TN}+\text{FP})(\text{TN}+\text{FN})}},
\]
being equal to $+1$ for a perfect prediction, to $0$ for 
random classifications, and to $-1$ for wrong predictions.}
, or by computing the balance point, namely the 
point for which $\mathrm{TPR = 1-FPR}$. In this work we use the latter approach as the default probability threshold, emphasizing that we compared this value to those computed with the $F_1$-score and the Matthews coefficient, finding similar results between these thresholds and the balance point, which in turn mean that there are no significant differences for the performance metrics that we present in \cref{sub:Random forest classification performance}.
\begin{figure*}
  \begin{center}
    \includegraphics[width=0.98\textwidth]{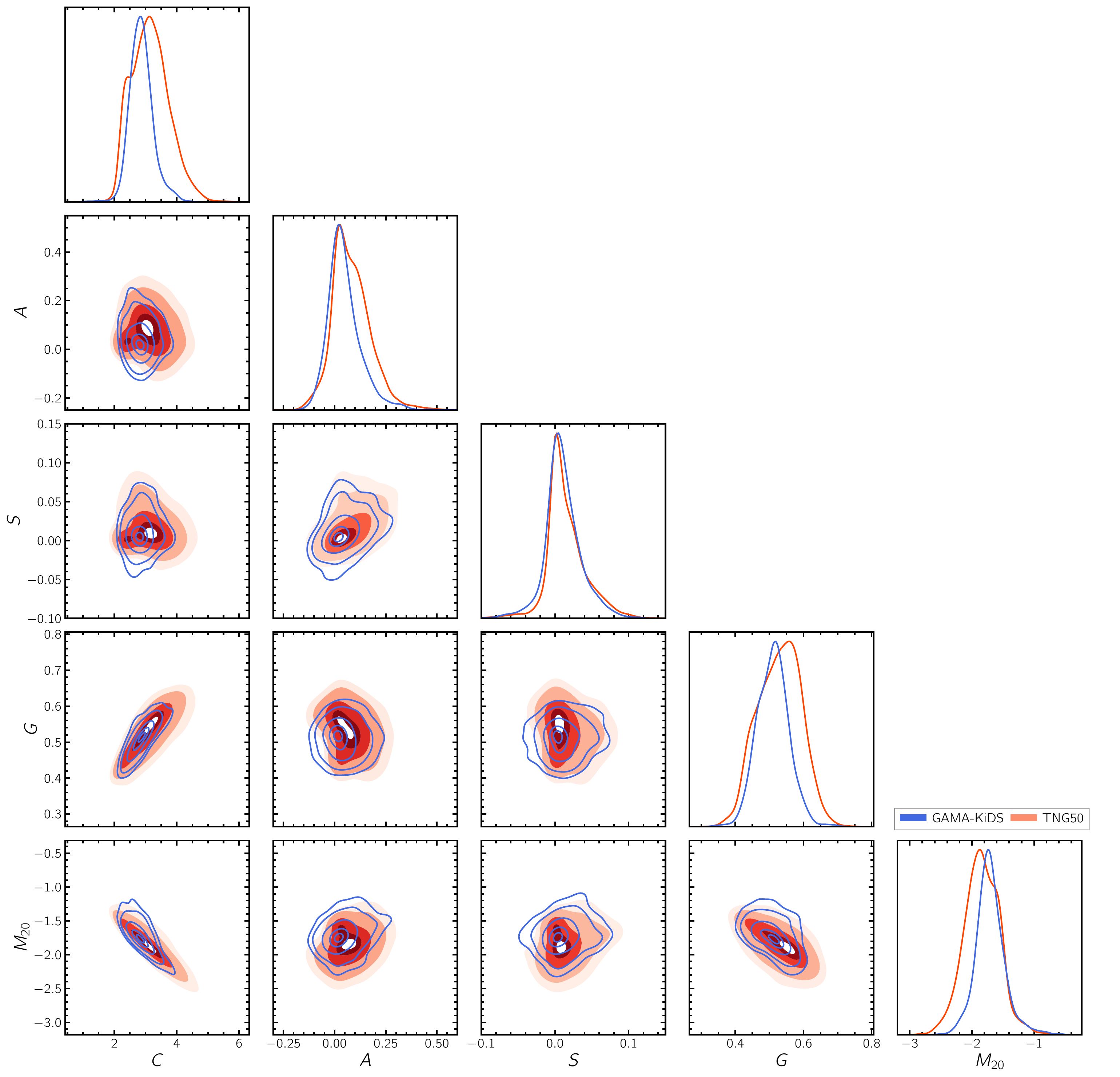}
  \end{center}
  \caption{Pairwise plots of $r$-band morphological parameters from 
    the observational GAMA-KiDS (blue) and simulated TNG50 (red) samples, with univariate distributions shown on the diagonal. The bi-dimensional kernel density estimates are shown with contours at $\{0.1, 0.2, 0.5, 0.8, 0.95\}$. This plot indicates that there is good overall agreement between the morphologies of these samples.}
\label{fig:pairs_obs_Sim_comb}
  \end{figure*}

\begin{figure*}
  \begin{center}
    \includegraphics[width=0.88\textwidth]{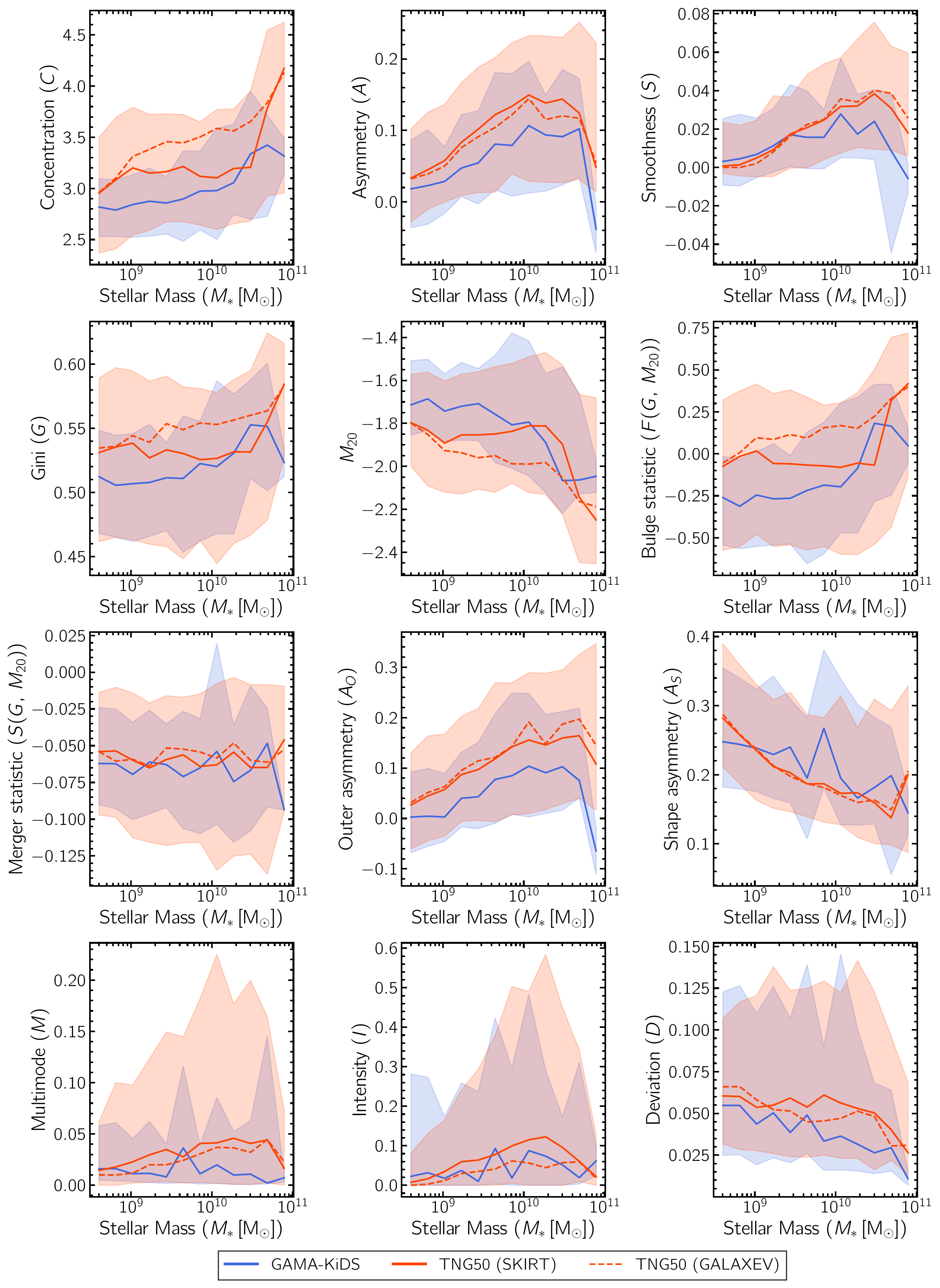}
  \end{center}
  \caption{Median trends as a function of stellar mass for 
    several $r$-band  morphological parameters.
The solid red and blue lines indicate the simulated (SKIRT pipeline) 
and observational results, respectively; 
the dashed line corresponds to synthetic images generated 
without the effects of a dust
distribution (GALAXEV pipeline). This figure again shows that there is good agreement between theory and observations, with the 
median values (at fixed stellar mass) of all morphological parameters in TNG50 lying within 
1$\sigma$ of the observational trends; it also shows that the simulated galaxies are more concentrated and asymmetric than their observational counterparts.}
\label{fig:medians}
  \end{figure*}
\begin{figure*}
  \begin{center}
    \includegraphics[width=0.98\textwidth]{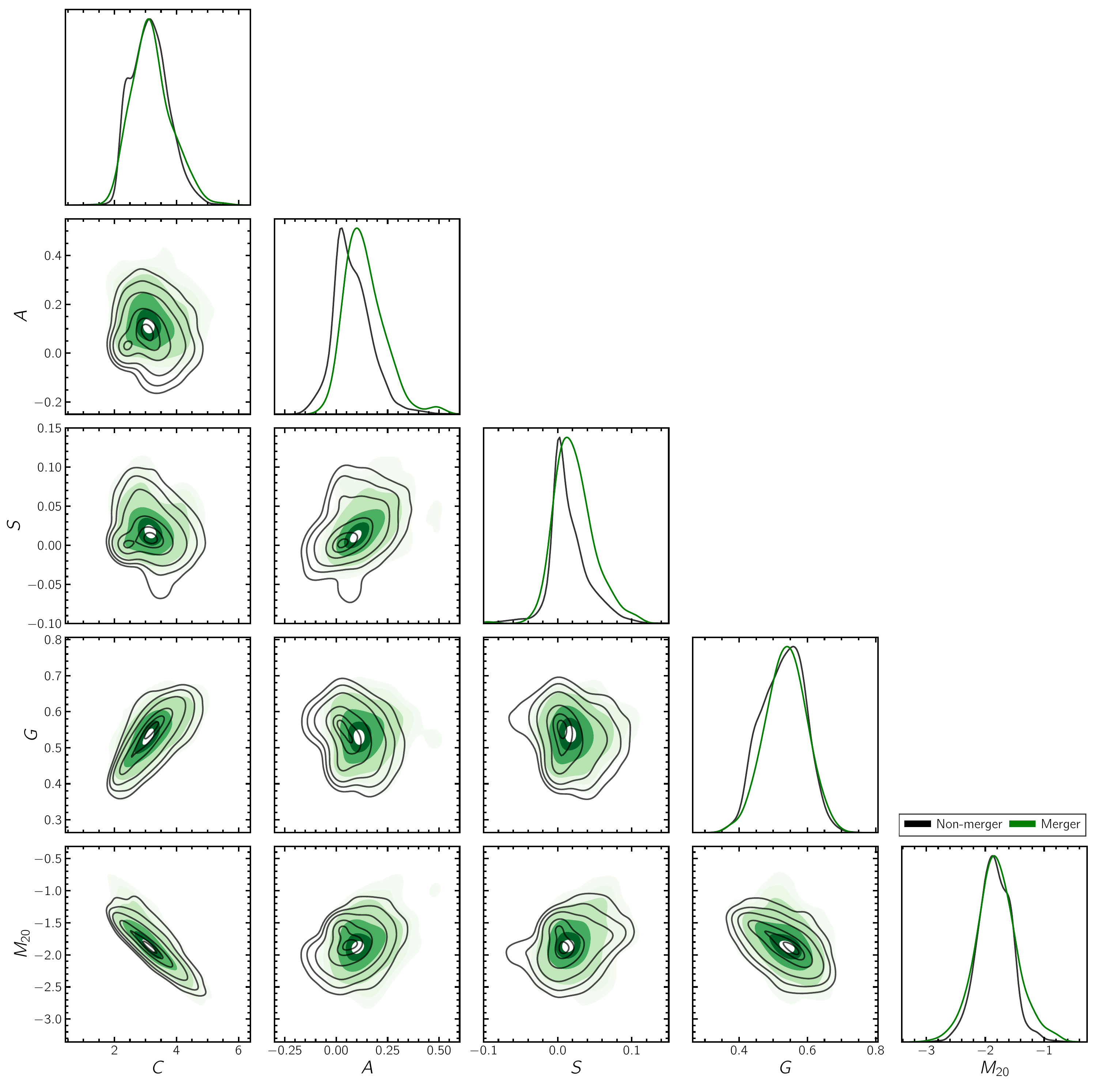}
  \end{center}
  \caption{Pairwise distributions of $r$-band morphological parameters for the non-merging
  population (black) against the distributions of mergers
  (green), with contours located at 
  $\{0.05,0.1,0.2,0.5,0.8,0.95\}$. This
figure demonstrates that simulated merging and non-merging systems
exhibit a high degree of overlap in their morphologies, 
occupying similar regions in
parameter space; it also indicates that the distributions of the merging population tend to have higher distribution values,
which is more noticeable for the asymmetry parameter.}
\label{fig:pairs_merg_nonmerg}
  \end{figure*}
\begin{figure*}
  \begin{center}
    \includegraphics[width=0.88\textwidth]{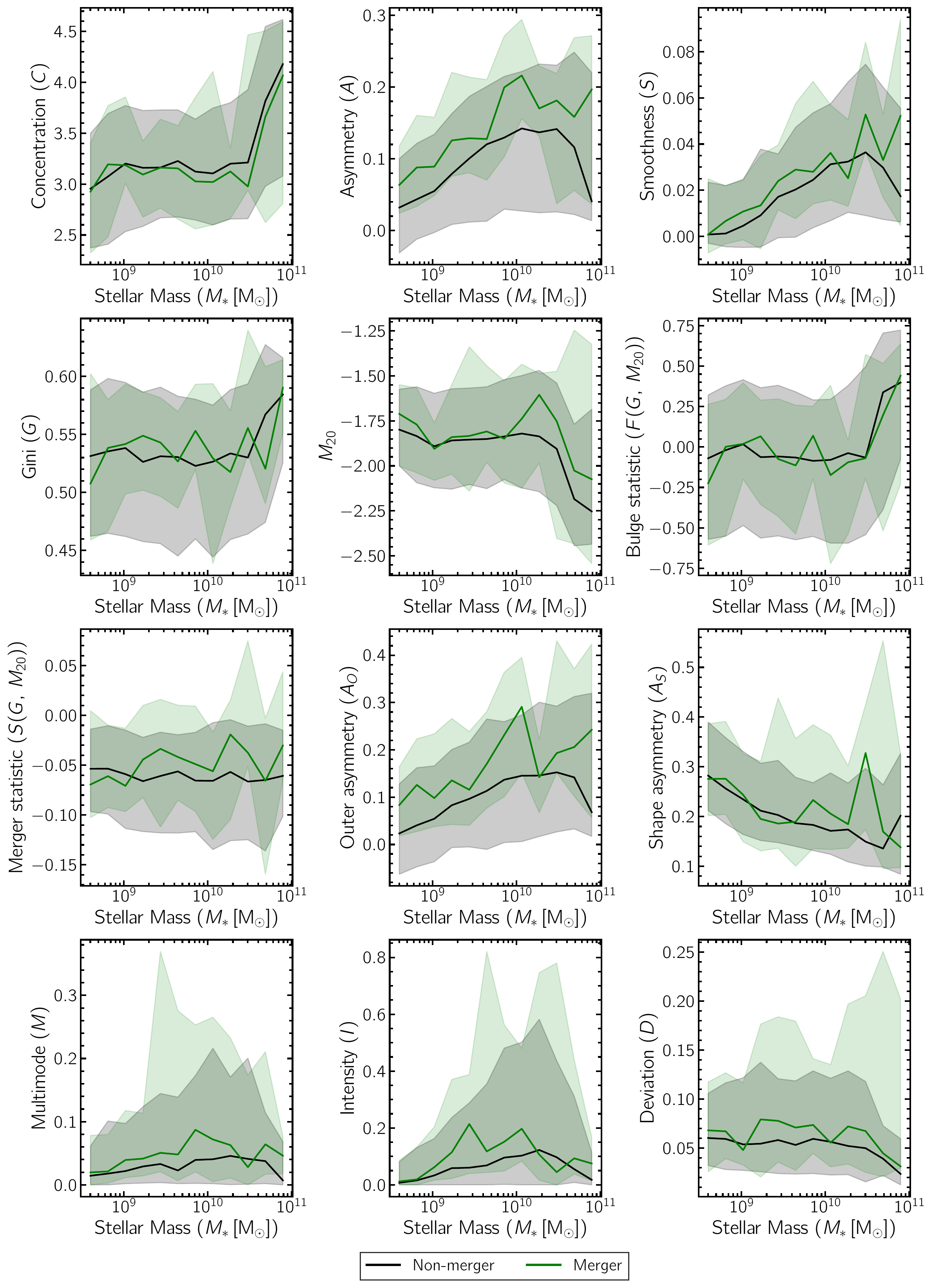}
  \end{center}
  \caption{Median trends as a function of stellar mass for 
    several $r$-band morphological parameters.
The solid black and green lines indicate the low-mass non-merging and 
merging 
TNG50 galaxy populations, respectively. Although merging systems tend to be more asymmetric at all stellar masses, this plot shows that the morphologies of 
these two samples are highly comparable, particularly at the low mass end.}
\label{fig:medians_merg_nonmerg}
  \end{figure*}
 \begin{figure*}
  \begin{center}
    \includegraphics[width=0.97\textwidth]{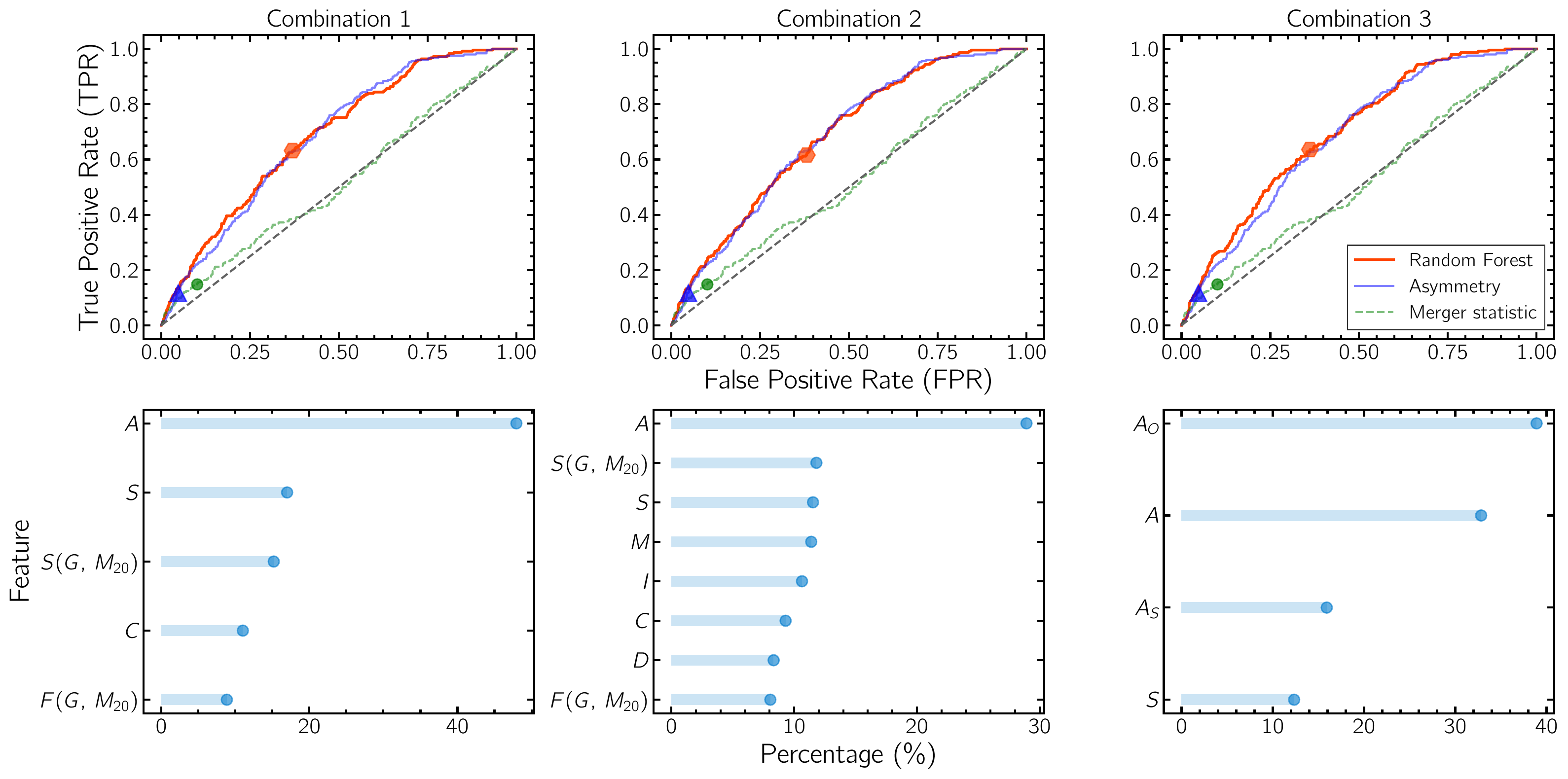}
  \end{center}
  \caption{Top row: ROC curves for random forest models trained on the
    full dataset using the combinations in listed in
\cref{tab:comb_features}; asymmetry and merger statistic
lines indicate merger
  selections using only one of these parameters. The hexagons, 
  triangles
  and circles indicate default threshold values for the classifier
  (balance point), asymmetry parameter ($0.25$) 
  and merger statistic ($0.0$) selections. Bottom row: feature
importance for each model corresponding to the random forest ROC curves above. 
The vertical axis indicates the name of the
corresponding attributes, while the horizontal axis denotes the
percentage, i.e., the parameters' contribution to classification
decisions. In all cases the asymmetry or outer asymmetry 
parameters had the highest
weight on decisions, with the merger statistic, smoothness or asymmetry 
parameter (combination 3) 
ranking second in their respective 
combination.}
  \label{fig:roc_fimp}
  \end{figure*}

\section{Results}%
  \label{sec:results}
 \subsection{The morphologies of TNG50 and GAMA-KiDS galaxies}%
   \label{sub:The morphologies of TNG50 galaxies}
\cref{fig:pairs_obs_Sim_comb} 
shows pairwise plots of various morphological 
parameters measured by \textsf{statmorph} for our simulated 
(TNG50, red shaded regions) and observational 
(KiDS, blue contours) galaxy samples. 
In the former case, we only show 
measurements for a single projection 
(onto the $xy$-plane).
The plots in the lower triangle of \cref{fig:pairs_obs_Sim_comb}
show the joint 
distributions of the various morphological parameters, while the 
plots on 
the diagonal 
show their univariate distributions.
As can be seen, there is reasonable 
agreement between the
morphologies of simulated galaxies and those of the 
real observational sample.
However, \cref{fig:pairs_obs_Sim_comb} also reveals some differences 
between the two samples. For instance, the concentrations and Gini 
coefficients of TNG50 galaxies peak at slightly higher values than 
their observational counterparts, while the distribution of the 
$M_{20}$ 
parameter reaches a maximum at a slightly lower value. These trends 
indicate that TNG50 galaxies tend to be slightly 
more concentrated objects 
than 
real galaxies of similar stellar mass. On the other hand, the
asymmetry 
distribution reaches a peak very close to 
zero for both 
samples, but displays 
a tail toward higher values for TNG50 galaxies, 
indicating that these objects tend to be slightly more
asymmetric than their observational counterparts.

\cref{fig:medians} 
shows median trends as a function of stellar mass for all 
morphological parameters
considered. The red
solid line 
corresponds to
galaxies modelled following our full radiative transfer pipeline, 
while the
dashed one was obtained using simpler models without 
including the effects of 
a dust distribution; similarly, the blue solid line indicates 
the median trend for our 
GAMA-KiDS sample. The blue and red shaded regions denote 
the corresponding 16th to 84th percentile range, at a fixed 
stellar mass, for the observational and simulated 
(dust effects included) samples, respectively. 
These figures confirm
the overall morphological agreement between observations 
and simulations: all median trends from TNG50 lie 
within the $1\sigma$
scatter of the observational
measurements. However, closer inspection of \cref{fig:medians} 
shows again that TNG50 galaxies 
tend to be slightly more concentrated and asymmetric than 
their observational counterparts.

Finally, \cref{fig:medians} corroborates that simulated 
galaxies modelled with 
dust effects have morphologies
better aligned with observations, since dust attenuation tends to 
reduce the brightness of the
central regions, where higher concentrations of gas and dust are 
typically encountered in the simulation.
This effect, however, is not enough to bring the 
concentrations of low-mass galaxies into full agreement 
with observations.
\subsection{The morphologies of true mergers and non-mergers}%
  \label{sub:The morphologies of intrinsic mergers}
\cref{fig:pairs_merg_nonmerg} 
shows morphology distributions for the merging sample 
(green shaded regions)
defined in \cref{sub:Merger identification} as well as for the 
non-merging TNG50 population (black contours). As can be seen, the
morphologies of these two groups do not differ significantly from
each other, i.e. they occupy similar regions in
parameter space, as previously pointed out
for the case of galaxies at higher redshifts \citep{Snyder2019}.
The morphological similarity between mergers and non-mergers can also be seen in \cref{fig:medians_merg_nonmerg}, which shows 
median trends (solid lines) and $1\sigma$ 
scatter regions (shaded zones) 
for the relevant morphological parameters as a function of stellar mass. 
Nevertheless, the asymmetry and outer asymmetry parameters 
reveal that, 
although there is significant morphological overlap
between mergers and non-mergers, the former are more asymmetric 
than the latter. In the next sections, we exploit these differences in 
order to train an image-based galaxy merger classifier.

\subsection{Classifying mergers with random forests}%
\label{sub:Random forest classification performance}
\subsubsection{Performance and feature importance}
In this section we present results about the merger classification
of the mock-catalog as
detailed in \cref{sub:Random forest classification}. 
The hyper-parameter tuning and cross-validation procedures performed
on the training sets result in
the best trained models for each of the feature combinations listed 
in \cref{tab:comb_features}. Each of these models was then
applied to the corresponding test set, which had ${\approx}4389$
non-merging objects and
${\approx}250$ merging galaxies, with stellar masses
$8.5\leqslant\log \qtty( M_\ast / \mathrm{M}_\odot)\leqslant11$.
We found that all models yield similar classifications 
independently of the feature 
combination used: ${\approx}179$ and $2425$ objects were correctly
classified as mergers and non-mergers, respectively; $1964$
non-merging galaxies were misclassified as mergers, 
and $71$ mergers were
misclassified as non-mergers.
These findings are translated, for the merger class, 
into an average purity and completeness of ${\approx}8.4\%$ and
${\approx}72\%$, respectively.\footnote{A random classifier
(without predictive power) would have a purity of 5.4\%, 
equal to the overall
merger fraction.}

The upper panels of \cref{fig:roc_fimp} show the ROC curve (see \cref{sub:Random forest classification}) 
for the trained models using combinations 1--3 from
\cref{tab:comb_features}. For reference, we also show the 
ROC curves that
would be obtained by using only 
the asymmetry parameter or the Gini--$M_{20}$ merger statistic to
select
merging galaxies.
A perfect classifier lies at the upper-left corner
and has $\text{TPR}=1$ and
$\text{FPR}=0$, while a model that goes in diagonal from $(0,0)$ to 
$(1,1)$ has no predictive power. Our models have loci between these
two regions, with an area under the ROC curve of ${\approx}0.7$,
indicating 
that they are moderately adequate
classifiers.  

The lower panels of 
\cref{fig:roc_fimp}. show, also for the model combinations 1--3, 
input features and their relative importance to the
classification decisions. In this context, feature importance is
defined as the mean decrease in impurity achieved by each variable
at all relevant nodes in the random forest.
In all cases the asymmetry or outer asymmetry parameters 
had the highest weight ($\sim30-50\%$) on the classifier
decisions, followed in second place by smoothness parameter ($15\%)$ or 
the merger statistic ($12\%)$.
None of the other parameters had a feature importance above 15\%. 

\subsubsection{Default random forest model}%
  \label{sub:Default_rf_model}
The models presented above have similar performance regarding
purity and completeness values. These results are consistent with
the work by \citet{Snyder2019}, who designed random forest
models for high-mass galaxies at different redshifts from 
the original
Illustris simulation. For their lowest redshift sample at $z=0.5$,
they obtained purity values of up to $10\%$, with completeness at
roughly $70\%$. Similarly, the metrics produced by our models are 
consistent
with those found by \citet{Bignone2016} for the Illustris simulation.
Specifically, they studied the morphologies of a galaxy sample at $z=0$ 
with $M_\ast > 10^{10}\,\text{M}_\odot$ and subsequently used the 
Gini--$M_{20}$ criterion from \citet{Lotz2004} to identify 
galaxy mergers. Their results, as a function of the time $t$ elapsed 
since the last merger, show that for $t\sim1$ Gyr 
the purity metric is around $5\%$ for cases with 
$\mu > 1/10$ and equal to $9\%$ for $\mu > 1/4$.
Throughout the 
rest of this paper we set
the RUS+RF model trained with combination 
one in \cref{tab:comb_features} as
our default random forest model. This decision is mainly based on
the robustness of classification performance 
for different sets of
features, but also because the 
morphological parameters included in that combination are 
widely used 
in the literature (e.g. \citealt{Lotz2011}, and references therein)
to identify both major and minor mergers. 
In \cref{sub:The merger incidence of GAMA-KiDS observations} 
we use this model to estimate the galaxy merger fraction 
in the TNG50 simulation, which we compare 
to the intrinsic merger fraction 
(i.e. computed with the merger trees), as well as to estimate the 
galaxy merger fraction in the real Universe by applying our classifier to 
GAMA-KiDS
observations.

\begin{figure*}
  \begin{center}
    \includegraphics[width=0.49\textwidth]{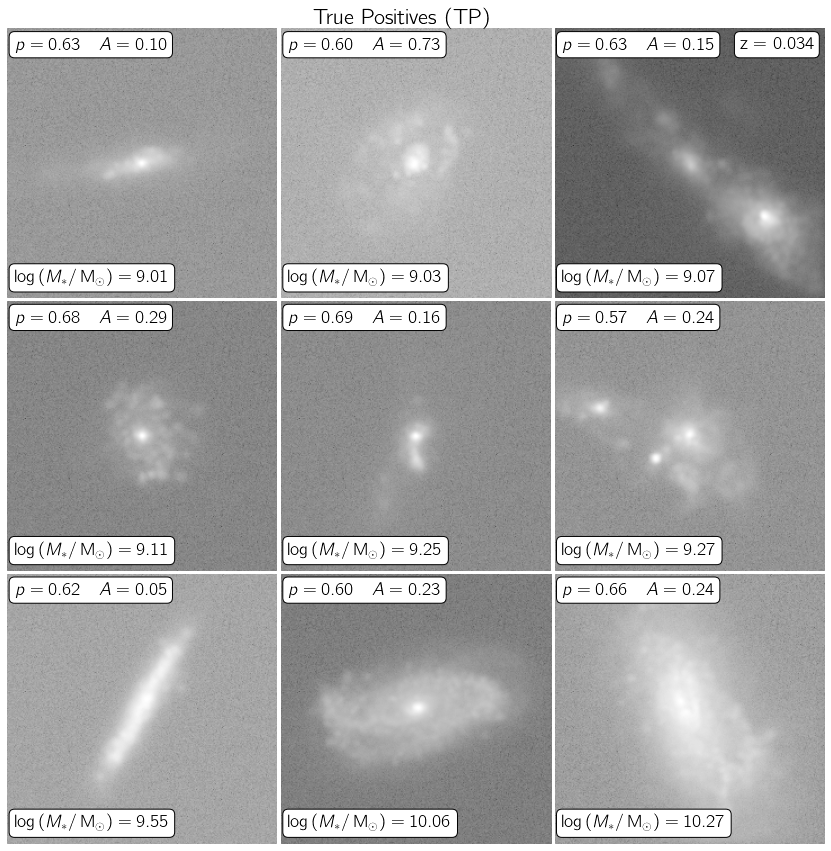}
    \includegraphics[width=0.49\textwidth]{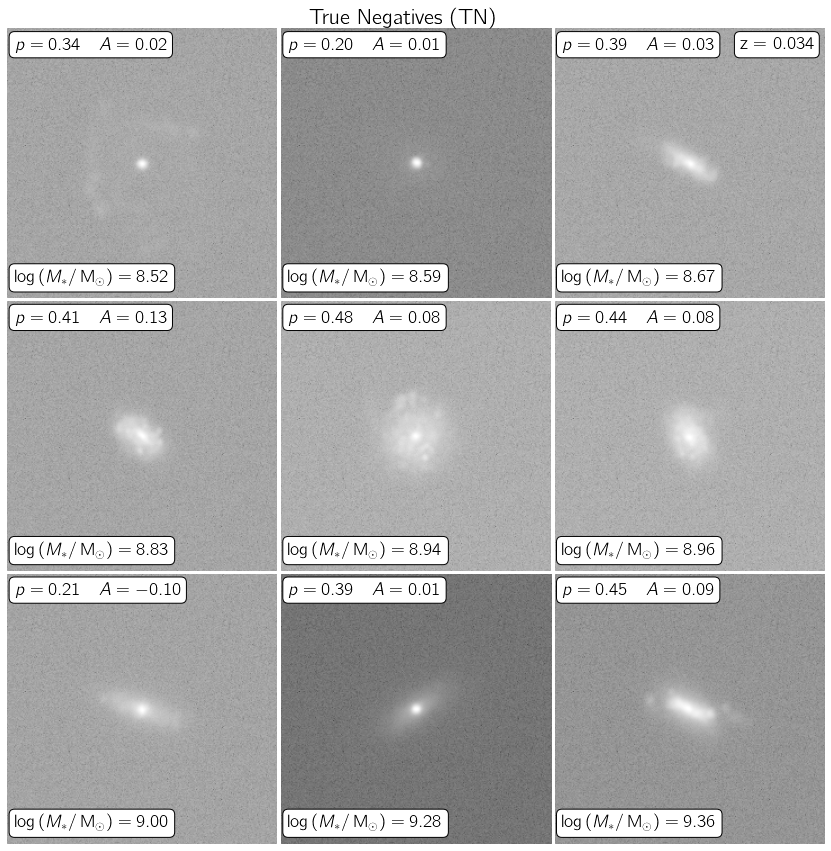}
    \includegraphics[width=0.49\textwidth]{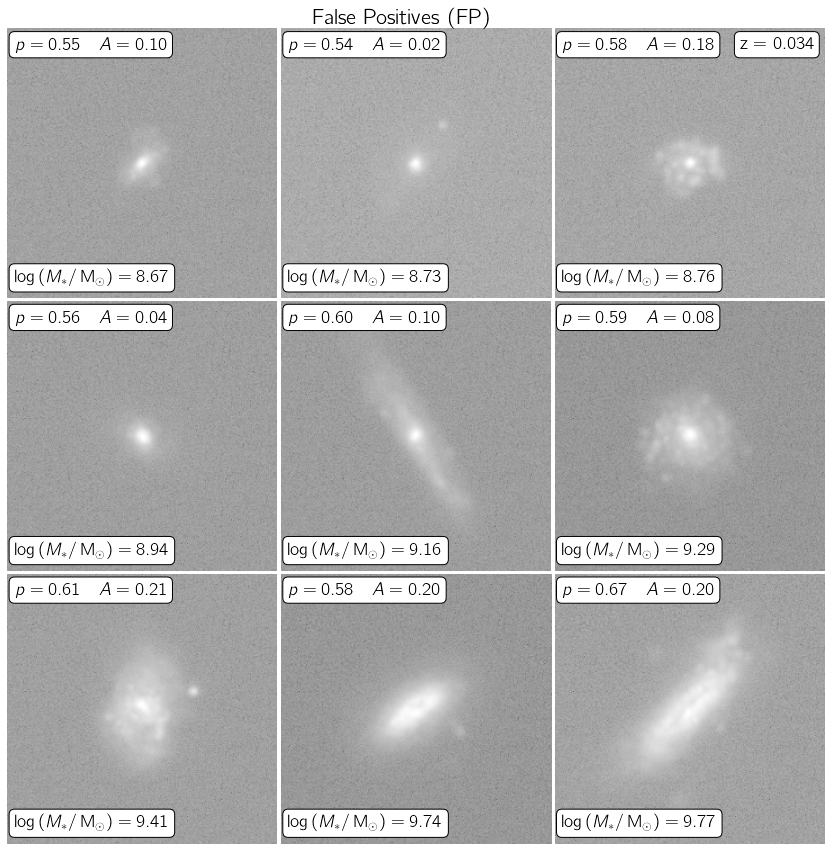}
    \includegraphics[width=0.49\textwidth]{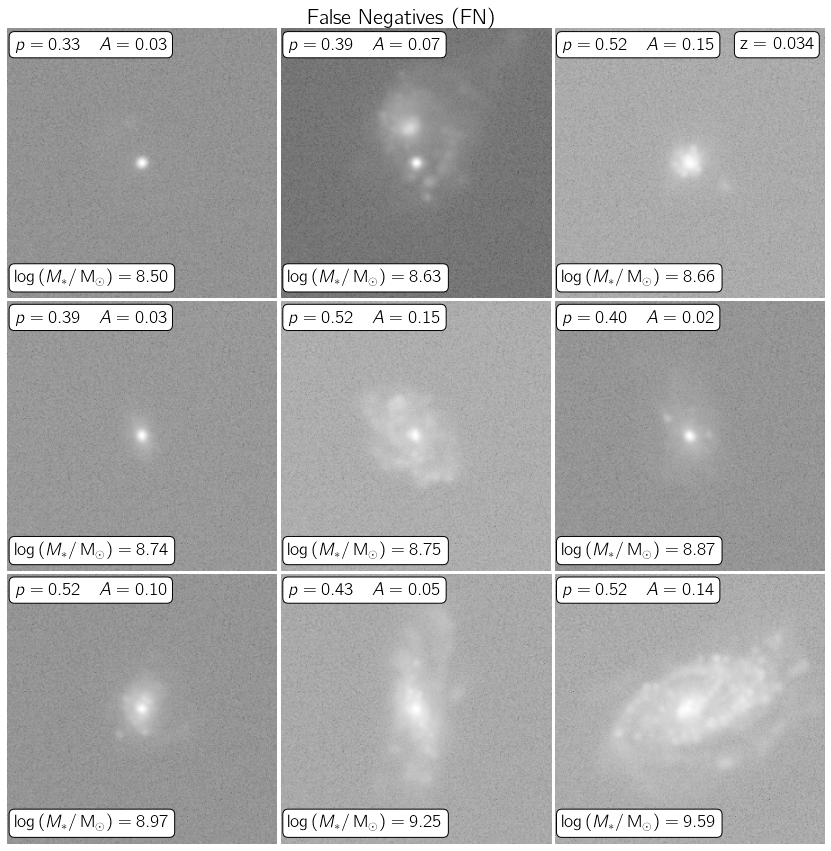}
  \end{center}
  \caption{Classification result examples made by the random forest model.
  The upper-left block of figures shows true positives (true mergers
selected) while the upper-right block shows true negatives (non-mergers
rejected). These objects typically exhibit the expected 
attributes: true
positives (mergers) are
perturbed and often have companions, whereas true negatives
(non-mergers) tend to be
isolated and unperturbed. 
Similarly, the lower-left block of figures shows false positives (non-mergers
selected) while the lower-right part shows false negatives (true mergers
rejected). False positives might arise from mergers taking place
outside the window detection period or by having similar
morphologies to some true mergers; false negatives
could emerge from minor mergers failing to trigger morphological disturbances 
at the time of detection.
In all cases the upper text labels indicate the probability value assigned 
to that object by the random forest, as well as its 
corresponding asymmetry value.}
  \label{fig:tp_tn_fp_fn}
  \end{figure*}

\subsubsection{Classification result examples}%
  \label{sub:Classification result examples}
In \cref{fig:tp_tn_fp_fn} we present examples of simulated
galaxies that have been classified by our default random forest 
model and were
categorised as true positives and true negatives
as well as false
positives and false negatives. 
Note that these objects are sorted
in ascending order according to stellar mass.
As can be seen in the upper row from 
\cref{fig:tp_tn_fp_fn}, most true mergers exhibit
clear signs of interaction, such as asymmetric structures and
neighbouring or overlapping companions. Likewise, 
most non-mergers
do not have significantly perturbed morphologies 
and look relatively isolated. This is particularly 
noticeable for low-mass
objects.

Similarly, the lower row in \cref{fig:tp_tn_fp_fn} 
shows examples of the failure modes of
the classifier. We found that some false positives had relatively
asymmetric structures but were not labelled as mergers in our
merger tree-based selection (see \cref{sub:Merger identification}). 
Thus, such cases
might arise from merging events taking place outside our window
detection period or from isolated galaxies that are morphologically 
similar to mergers
(see \cref{fig:pairs_merg_nonmerg,fig:medians_merg_nonmerg}). 
 
Conversely, some false negatives appear unperturbed so
that they are probably the result of minor mergers that did not
trigger perceptible morphological signatures. These cases are reminiscent of the findings by \citet{2021MNRAS.500.4937M}, who found 
that mergers induce
limited morphological changes in dwarf galaxies.
Thus, the lower part of \cref{fig:tp_tn_fp_fn} illustrates the most common 
challenges faced by the
models. On the one hand, merging events are rare, which reduces the 
number of class examples and statistics for training; 
on the other hand, there is
a significant degree of similarity between mergers and
non-mergers, which
explains most of the false positives.   
 \begin{figure*}
  \begin{center}
    \subfloat[\label{fig:merger_frac_rf}]{\includegraphics[width=0.45\textwidth]{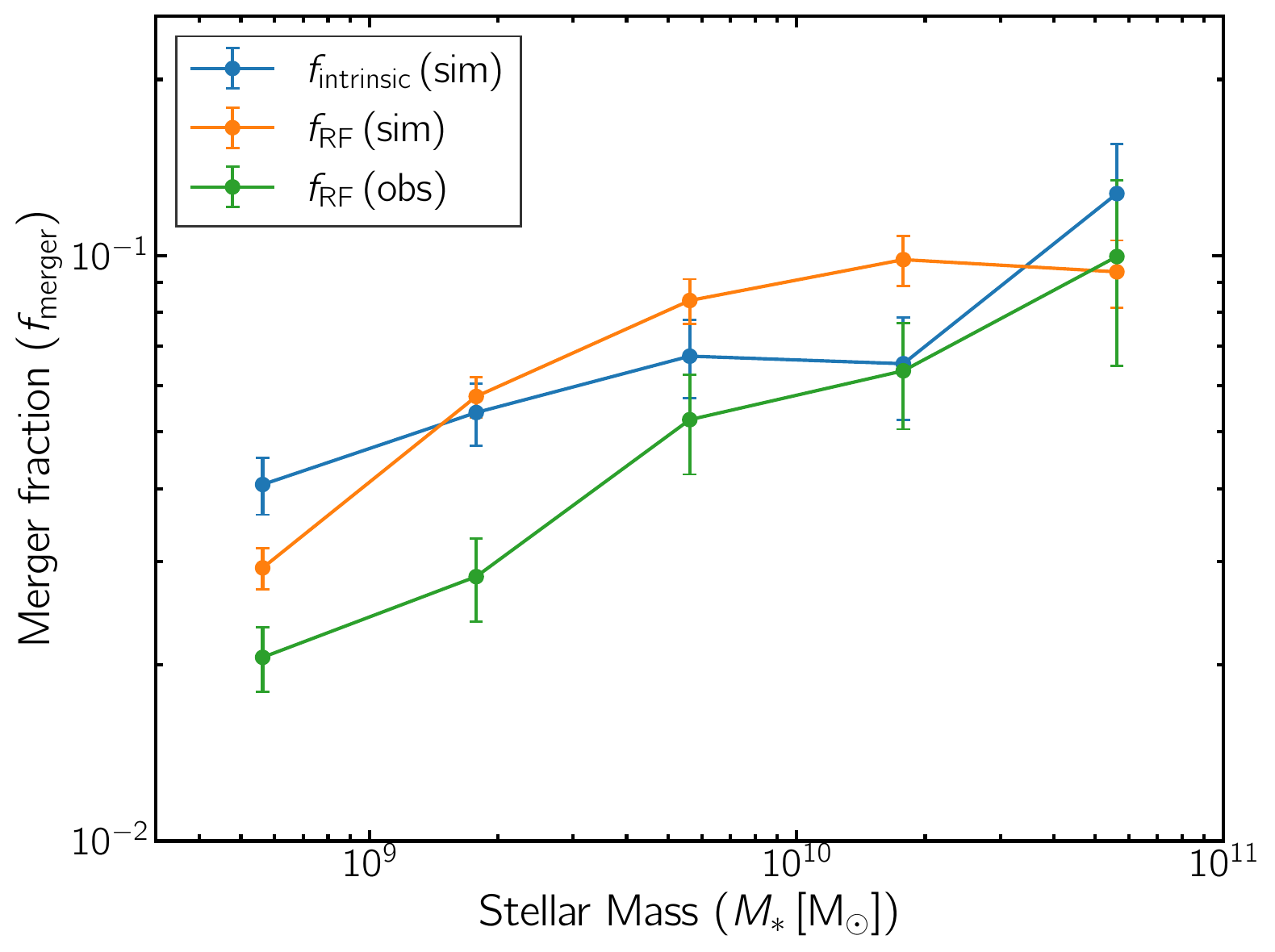}}
    \subfloat[\label{fig:merger_frac_asym}]{\includegraphics[width=0.45\textwidth]{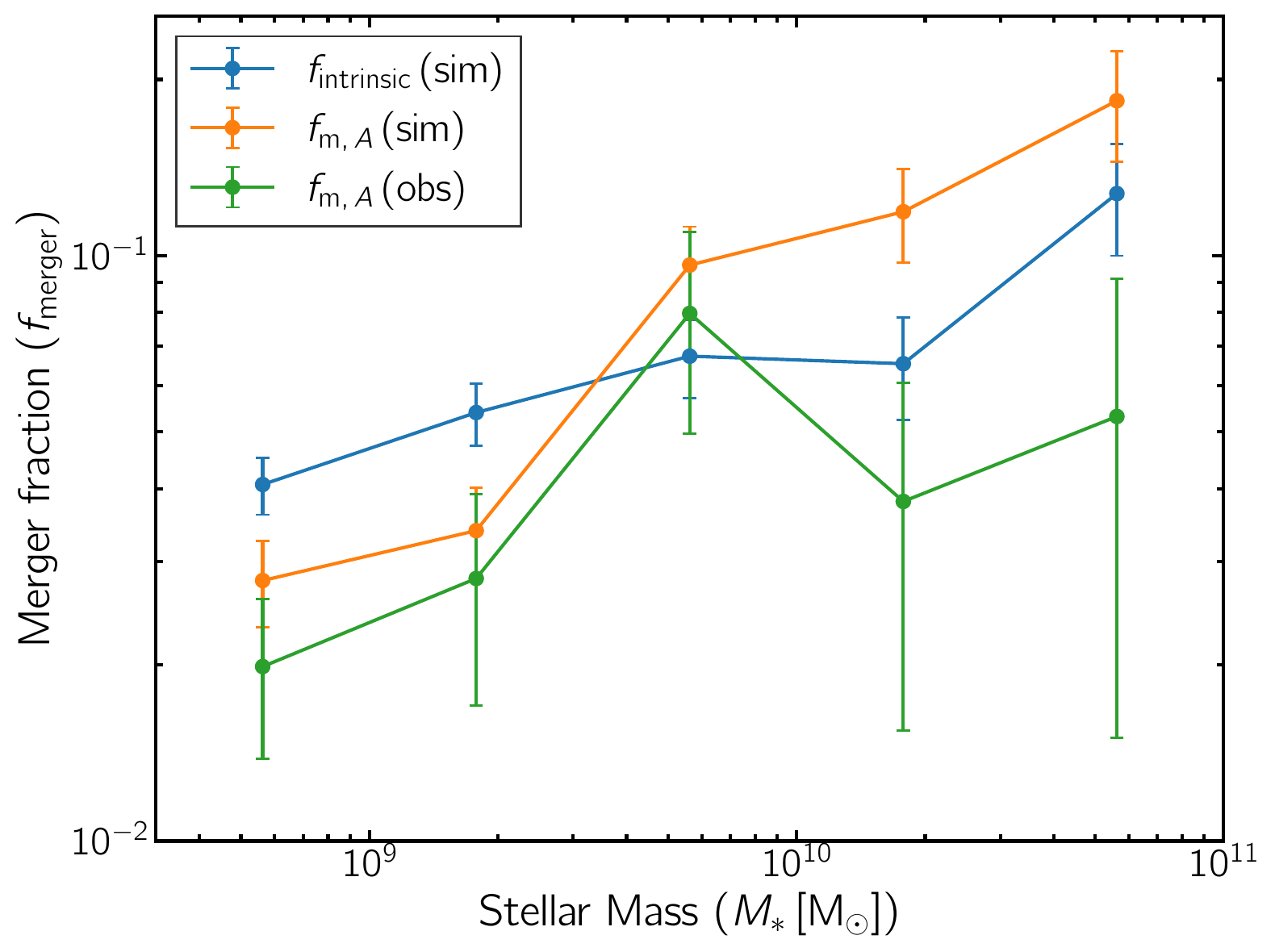}}
  \end{center}
  \caption{(a) Galaxy merger fraction as a function of 
    stellar mass as
  predicted by our random forest classifier for our simulated and 
  observational galaxy samples, 
  as indicated by the legend. (b) Mass-dependent merger fraction for
  the observational and simulated galaxy samples as predicted by
  an asymmetry-based criterion. In both panels, the solid blue 
  line represents the intrinsic merger fraction measured in 
  TNG50 using the merger trees. The error bars represent Poisson
uncertainty from the number of mergers in each mass bin. This figure 
shows that the merger fraction increases steadily as a function 
of stellar mass for all the approaches considered.}
  \label{fig:merger_frac_ab}
  \end{figure*}
    
\subsection{The merger incidence of GAMA-KiDS observations}%
  \label{sub:The merger incidence of GAMA-KiDS observations}
We used our default random forest model to estimate the merger
fraction in our observational 
$(z<0.05; 
8.5\leqslant\log \qtty( M_\ast /\mathrm{M}_\odot)\leqslant11)$
galaxy sample.
The input entries for this model consist of the
morphological measurements carried out on the GAMA-KiDS sample (see
\cref{sub:Morphological_measurements}), corresponding to 
combination one in
\cref{tab:comb_features}.

Following \citet{Snyder2019}, we estimate the merger fraction
$f_{\mathrm{ merger }}$ as
  \begin{equation}
    f_{\mathrm{ merger }}=\frac{ N_{\mathrm{ RF }} }{ N_{\mathrm{T}}
    }\frac{\mathrm{ PPV } }{ \mathrm{ TPR } }\expval{M/N},
    \label{eq:merger_frac_rf}
  \end{equation} 
  where $N_{\mathrm{ RF }}$ and $N_{\mathrm{T}}$ 
  are the number of galaxies
selected by the forest and the total number of galaxies,
respectively; the factor $\expval{M/N}$ is the average total number 
of simulated mergers divided by the number of galaxies with at
least one such merger, which in our case is approximately
equal to one; the term $\mathrm{ PPV }/\mathrm{ TPR }$ is a 
corrective factor that takes into account the completeness and 
purity of the default model, so that when used on novel data the 
best-guess merger fraction is obtained \citep{Snyder2019}.  
For comparison, we have determined the intrinsic merger fraction of
simulated galaxies by estimating, for each mass bin, the quantity
$N_{\mathrm{ merger }}/N_{\mathrm{T}}$, 
where $N_{\mathrm{merger}}$ is the
number of intrinsic mergers, obtained from the merger trees 
as described in
\cref{sub:Merger identification}.

\cref{fig:merger_frac_rf}
shows the estimated merger fraction, as a function of
stellar mass, for the simulated and observational samples. As
can be seen, both follow a qualitatively similar trend to the
intrinsic merger fraction, but show differences within a 
factor of ${\sim}2$,
as discussed below. The error bars in the estimated merger
fractions are given by Poisson statistics 
from the number of mergers in
each bin.
The fact that 
the random forest classifier predicts a lower merger fraction for 
observed galaxies than for simulated ones is perhaps not 
surprising by 
noting that the most important feature for the classifications 
is the 
asymmetry parameter 
(\cref{sub:Random forest classification performance}), 
and that observed galaxies have 
somewhat lower asymmetry values than simulated ones (\cref{sub:The morphologies of TNG50 galaxies}). 
A more robust result is that the galaxy merger fraction is an increasing
function of stellar mass for all the cases considered.

These findings are in contrast with 
the major merger fraction estimate by
\citet{Casteels2014}, who found a decreasing merger fraction 
within the 
stellar mass range 
$9.0\leq\log \qtty( M_\ast / \mathrm{M}_\odot)\leq9.5$, and a 
roughly
constant trend above that mass range, while all of our estimates 
indicate that the merger fraction increases steadily as a function
of stellar mass.
This qualitative difference is puzzling, 
considering that the estimate by \citealp{Casteels2014} was performed
on a similar observational galaxy sample, using the asymmetry 
parameter  
to estimate the fraction of asymmetric galaxies and subsequently
the major merger fraction as a function of stellar mass. 

For comparison, we have applied an asymmetry-based criterion to our
samples to identify highly asymmetric galaxies. 
These objects were selected as those for which
$A>0.25$. In the following step we computed the PPV and TPR for
these predictions, and we applied a modified version of 
\cref{eq:merger_frac_rf} to estimate the merger 
fraction
derived from the asymmetry criterion. We note that
$\text{PPV}\approx8.5\%$ and $\text{TPR}\approx7.5\%$ for 
this classification,
which represents a purity of the same order as that of our RF models 
but with a completeness
that is
considerably smaller.

\cref{fig:merger_frac_asym} shows a comparison between the
intrinsic merger fraction from the simulation and the
asymmetry-based merger fraction, $f_{\mathrm{m},\,A}$, 
for both simulation and observations.
As can be seen, 
such estimates qualitatively follow the trend of the 
intrinsic merger fraction: both are increasing functions of 
stellar mass. However,
$f_{\mathrm{m},\,A}(\mathrm{obs})$ is smaller than the other 
two fractions by a factor 
of ${\sim}2$, which again reflects the fact that the 
asymmetry parameter tends 
to be lower for our observational sample than for our simulated one. 

\section{Discussion}%
\label{sec:Discussion}

Using the state-of-the-art TNG50 cosmological simulation and 
KiDS observations, we have studied the optical morphologies 
of galaxies at low redshift ($z < 0.05$) over a wide range 
of stellar masses ($8.5 < \log_{10}(M_\ast/\mathrm{M}_\odot) < 11$). 
The goal of this analysis has been threefold: (i) to carry out an
`apples-to-apples' comparison between the optical morphologies of
TNG50 and KiDS galaxies, allowing us to identify possible weaknesses
in the IllustrisTNG galaxy formation model at unprecedentedly 
high mass resolution (16 times better than TNG100); 
(ii) combining morphological measurements of the 
simulated galaxies with information from the merger trees, 
to train and evaluate the performance of an algorithm for identifying merging 
galaxies based on morphological diagnostics alone; and (iii)
to apply this simulation-trained algorithm to observations 
in order to estimate the galaxy merger fraction in the real Universe.

The first step for carrying out this work was to prepare the
observational data set, shown in 
\cref{fig:gamakidstiles,fig:z_vs_mass}, which consisted in selecting
galaxies from the GAMA catalogues satisfying 
$8.5 \leqslant \log_{10}(M_\ast/\mathrm{M}_\odot) \leqslant 11$ 
and $z < 0.05$, 
and extracting their corresponding `cutouts' from KiDS 
mosaic images. Similarly, we prepared a simulation data set by 
selecting TNG50 galaxies from snapshot 96
(corresponding to $z = 0.034$, close to the median redshift of 
the observational sample) also satisfying 
$8.5 \leqslant \log_{10}(M_\ast/\mathrm{M}_\odot) \leqslant 11$, 
and then 
generating synthetic images for all simulated galaxies 
(including the effects of dust attenuation and scattering, and for 
three different projections) designed to match the KiDS data set.
\cref{fig:comp_images} shows idealized, composite ($g,r,i$ bands) 
images for some of our simulated galaxies, while 
\cref{fig:sim_vs_obs_synth_images} shows the corresponding 
$r$-band images 
after including realism (convolution with a PSF and noise modelling), 
along with some example galaxies from the observational sample.

After preparing the observational and simulated data sets, we 
performed source segmentation and deblending on each image in 
order to isolate the galaxy of interest and remove unwanted 
or contaminating sources, as illustrated in \cref{fig:deblending}. 
We then measured various morphological diagnostics in the $r$-band 
for galaxies from 
both data sets using the same code (\textsf{statmorph}), 
which represents a robust, quantitative comparison between 
theory and observations. This comparison showed good overall
agreement between TNG50 and KiDS galaxies, with the median trend 
as a function of stellar mass for TNG50 galaxies lying within
$\sim$1$\sigma$ of the observational distribution for every
morphological parameter considered 
(\cref{fig:pairs_obs_Sim_comb,fig:medians}). 
However, TNG50 galaxies tend to be slightly more concentrated and 
asymmetric than their observational counterparts, and 
show wider distributions for most parameters.

Interestingly, using the TNG100 simulation, \citet{Rodriguez-Gomez2019} also found that 
some IllustrisTNG galaxies are more concentrated compared to 
their observational counterparts from the 
Pan-STARRS $3\pi$ Survey \citep{Chambers2016}. 
However, this discrepancy was observed at higher masses, 
$M_{\ast} \sim 10^{11} \, \Msun$, and was attributed to the
implementation details of the active galactic nuclei (AGN) feedback -- specifically, 
it was argued that the spherical region over 
which energy and momentum are injected by the AGN might be 
too large, and therefore ineffective at small radii. 
In the present work we reach much lower masses than those 
achievable in TNG100 (by a factor of 16) and find that the 
discrepancy pointed out by \citet{Rodriguez-Gomez2019} 
reappears at $M_{\ast} \sim 10^{9} \, \Msun$. 
Despite the different stellar masses, it is possible that the 
reason for the higher concentrations of TNG50 galaxies at 
$M_{\ast} \sim 10^{9} \, \Msun$ relative to observations 
is essentially the same as that for TNG100 galaxies at 
$M_{\ast} \sim 10^{11} \, \Msun$, namely, inefficient AGN 
feedback at 
the smallest radii. While the AGN feedback implementation operates in 
very different modes in such different mass ranges 
(`thermal' versus `kinetic'; \citealt{Weinberger2017a}), 
the size of the `injection region' is determined by the 
same prescription in both feedback modes 
(a sphere enclosing an approximately fixed number of gas cells). It 
will be interesting and important to explore in future galaxy 
formation models whether reducing the size of the injection region 
for AGN feedback produces galaxies with concentrations in better 
agreement with observations. We note, however, that TNG50 produces deficits in the 
star formation density on small scales that agrees well with observations 
\citep{10.1093/mnras/stab2131}

On the other hand, the slightly higher asymmetries 
of TNG50 galaxies compared to observations could 
simply be a matter of resolution. Young stellar populations, 
in particular, are undersampled in hydrodynamic 
cosmological simulations, and manifest as bright clumps that 
become more noticeable in synthetic images produced with 
`bluer' broadband filters \citep{Torrey2015}. In principle, 
this issue could be mitigated by resampling the young 
stellar populations at a higher resolution 
\citep{Trayford2017}. However, such procedures would 
introduce additional complexity to our modelling and, 
importantly for our goal of characterising galaxy morphology, 
it is unclear what would be an appropriate spatial distribution 
for the resampled stellar populations. Therefore, we have 
adopted the simpler approach of smoothing the light 
contribution from every stellar particle using the same SPH-like
kernel, regardless of the age of the stellar population. 
A related issue is that the outskirts of simulated galaxies are subject 
to particle noise, which could further contribute to overestimating 
the asymmetry parameter. It seems plausible that the asymmetries of 
simulated galaxies will automatically become more realistic 
with improved resolution, without the need to make
substantial changes to the galaxy formation model.

Having compared the optical morphologies of TNG50 galaxies to those 
from KiDS observations, we proceeded to compare the morphologies 
of merging and non-merging galaxies in the simulation. 
In order to do this, we first defined a merger sample composed 
of simulated galaxies that experienced a major or minor merger 
(i.e. those with stellar mass ratio $\mu > 1/10$) within a time
window of approximately $\pm 0.5$ Gyr. We found that the 
morphology distributions of our merging and 
non-merging samples show a large degree of overlap, with 
the exception of the asymmetry-based statistics, as shown in 
\cref{fig:pairs_merg_nonmerg,fig:medians_merg_nonmerg}. However, 
despite such visually similar morphological distributions of 
our merging and non-meging samples, it is in principle possible that 
a combination of various morphological parameters would 
encode information about the merger histories of the galaxies that 
is unavailable when using the morphological parameters individually. 

To this end, we trained RFs using several combinations 
of morphological parameters of TNG50 galaxies as the 
model features, along with the merger label 
(0 or 1 for our non-merging and merging samples, respectively) 
as the ground truth, finding in all cases that 
the most important feature for identifying mergers is the 
asymmetry statistic. Therefore, the performance of our RF 
algorithm, usually quantified by the so-called ROC curve, 
is comparable to that of the more traditional method of 
selecting highly asymmetric galaxies, but is superior to a 
direct application of the Gini--$M_{20}$ merger statistic 
(\cref{fig:roc_fimp}). \cref{fig:tp_tn_fp_fn} shows some examples of 
the galaxy merger classifications 
(both successful and unsuccessful) returned by our RF algorithm.

The high importance of the asymmetry parameter might appear to be in tension 
with \citet{Snyder2019}, 
where the bulge indicators ($F(G,M_{20})$, concentration) had similar or 
greater importance than the asymmetry statistic, and the RFs clearly 
outperformed asymmetry alone. We attribute these differences to the distinct 
nature of the galaxies considered: massive galaxies 
($M_\ast > 10^{10}\,\text{M}_\odot$) at high 
redshifts in the case of \citet{Snyder2019}, and dwarf galaxies 
(mostly $M_\ast \lesssim 10^{10}\,\text{M}_\odot$) at low redshifts
in the present work. In fact, the RF classification models by \citet{2022arXiv220811164R} indicate that the asymmetry parameter is more significant for identifying mergers at low redshift than indicators of bulge strength (such as the concentration and Gini statistics), while the latter have a higher importance for high-redshift events. These findings help to reconcile our results with those of \citet{Snyder2019}. Another possible factor is the choice of broadband filters. The varying importance of different image-based merger diagnostics in different redshift and stellar mass ranges, as well as for different broadband filters, will be explored in upcoming work.

Finally, we applied our RFs to a test sample from the 
TNG50 simulation and to the observational sample, 
in order to estimate the galaxy merger fraction as a function 
of stellar mass in both simulations and observations 
(\cref{fig:merger_frac_rf}). In the case of the simulation, our RF 
was able to recover the `intrinsic' merger fraction 
(obtained directly from the merger trees) reasonably well 
(within a factor of $\sim$2).
When applied to KiDS observations, our RF returned a galaxy 
merger fraction that increases steadily with stellar mass, 
just like the intrinsic merger fraction in TNG50, 
although with a systematic offset of a factor of $\sim$2. 
For comparison, we repeated this experiment using the 
asymmetry statistic alone, separating mergers from non-mergers 
using a `standard' cut at $A = 0.25$ 
(\cref{fig:merger_frac_asym}). 
This yielded a steadily rising merger fraction in both 
simulations and observations, but again with a persistent 
offset between the two data sets, with the observational 
merger fraction lying a factor of $\sim$2--3 below the 
simulation trend. This offset probably reflects the fact that 
our simulated galaxies are slightly more asymmetric than their
observational counterparts. 

The results shown in \cref{fig:merger_frac_ab} imply that the merger fraction 
increases steadily with stellar mass, both when using the RF or the 
asymmetry parameter alone. These findings are qualitatively consistent with those of 
\citet{Besla2018}, who considered a low-redshift dwarf galaxy 
($0.013<z<0.0252$; $2\times10^{8}\,\text{M}_\odot<M_\ast<5\times10^{9}\,\text{M}_\odot$) sample from SDSS
to compute and compare the major pair fraction (the fraction of primary dwarf galaxies that have a 
secondary with a stellar mass ratio $\mu > 1/4$) with estimations
from the original Illustris simulation, also finding 
an increasing trend (their fig. 14). However, our results are
in stark contrast with those of \citet{Casteels2014}, 
who found a decreasing merger fraction over a comparable 
stellar mass range (their fig. 13), also using the asymmetry statistic as a 
merger indicator.

\section{Summary and outlook}
\label{sec:Summary}

We have carried out an `apples-to-apples' 
comparison between the optical morphologies of galaxies from 
the high-resolution, state-of-the-art TNG50 simulation and those 
of a comparable galaxy sample from KiDS observations. 
Overall, we have found good agreement between the simulated 
and observed data sets, which is remarkable considering that 
the IllustrisTNG galaxy formation model was not tuned to match
morphological observations. The TNG50 galaxies, however, are somewhat more 
concentrated and asymmetric than their observational counterparts. Using 
additional information from the simulation that is not available in observations 
-- namely, the merger trees -- we have trained a random forest algorithm 
to classify merging galaxies using image-based morphological 
diagnostics, and applied the random forest to observations in order 
to estimate the merger fraction in the real Universe. 
We found that the asymmetry statistic is the single most 
useful parameter for identifying galaxy mergers, 
at least in the mass and redshift regime we considered ($8.5\leqslant\log(M_\ast/\text{M}_\odot)\leqslant11$; $z<0.05$),
and that the merger fraction is a steadily increasing 
function of stellar mass for both the simulated and observational samples.

Currently, it is still challenging to 
precisely determine the merger fraction in observations, 
especially using galaxy morphology alone. However, 
we are approaching an era in which galaxy formation models 
will become so realistic that it will be possible to exploit 
subtle trends in morphological measurements 
-- such as the ones studied in this paper -- to infer properties 
of galaxies that are not directly observable in the real Universe, 
such as their merging histories or even the assembly histories of 
their host DM haloes. At the same time, our work highlights 
the importance of developing sophisticated tools to carry 
out robust comparisons between theory 
and observations, which will become indispensable in the 
upcoming years as both computational capacity and 
astronomical instruments continue to evolve. 


\section*{Acknowledgements}
We thank Gurtina Besla for useful comments and discussions.
VRG acknowledges support from UC MEXUS-CONACyT grant CN-19-154.
This work used the Extreme Science and Engineering Discovery Environment \citep[XSEDE;][]{Towns2014}, which is supported by NSF grant ACI-1548562. The XSEDE allocation TG-AST160043 utilized the Comet and Data Oasis resources provided by the San Diego Supercomputer Center.

The IllustrisTNG flagship simulations were run on the HazelHen Cray XC40
supercomputer at the High Performance Computing Center Stuttgart (HLRS) as 
part of project GCS-ILLU of the Gauss Centre for Supercomputing (GCS).
Ancillary and test runs of the project were also run on the compute cluster
operated by HITS, on the Stampede supercomputer at TACC/XSEDE 
(allocation AST140063), at the Hydra and Draco supercomputers at the 
Max Planck Computing and Data Facility, and on the 
MIT/Harvard computing facilities supported by FAS and MIT MKI. 

This research is based on observations made with ESO Telescopes at the 
La Silla Paranal Observatory under programme IDs 177.A-3016, 177.A-3017, 
177.A-3018 and 179.A-2004, and on data products produced by the 
KiDS consortium. The KiDS production team acknowledges support 
from: Deutsche Forschungsgemeinschaft, ERC, NOVA and NWO-M grants; Target;
the University of Padova, and the University Federico II (Naples). 

GAMA is a joint European-Australasian project based around a spectroscopic campaign using the Anglo-Australian Telescope. The GAMA input catalogue is based on data taken from the Sloan Digital Sky Survey and the UKIRT Infrared Deep Sky Survey. Complementary imaging of the GAMA regions is being obtained by a number of independent survey programmes including GALEX MIS, VST KiDS, VISTA VIKING, WISE, Herschel-ATLAS, GMRT and ASKAP providing UV to radio coverage. GAMA is funded by the STFC (UK), the ARC (Australia), the AAO, and the participating institutions. 

\section*{Data availability} 
The data from the IllustrisTNG simulations used in this work 
are publicly available at the website 
\href{https://www.tng-project.org}{https://www.tng-project.org}
\citep{Nelson2019}. The KiDS and GAMA data are available at the websites
\href{http://kids.strw.leidenuniv.nl/}{http://kids.strw.leidenuniv.nl/} and
\href{http://www.gama-survey.org/}{http://www.gama-survey.org/}.

\bibliographystyle{mnras}
\bibliography{bib}

\begin{thebibliography}{}
\makeatletter
\relax
\def\mn@urlcharsother{\let\do\@makeother \do\$\do\&\do\#\do\^\do\_\do\%\do\~}
\def\mn@doi{\begingroup\mn@urlcharsother \@ifnextchar [ {\mn@doi@}
  {\mn@doi@[]}}
\def\mn@doi@[#1]#2{\def\@tempa{#1}\ifx\@tempa\@empty \href
  {http://dx.doi.org/#2} {doi:#2}\else \href {http://dx.doi.org/#2} {#1}\fi
  \endgroup}
\def\mn@eprint#1#2{\mn@eprint@#1:#2::\@nil}
\def\mn@eprint@arXiv#1{\href {http://arxiv.org/abs/#1} {{\tt arXiv:#1}}}
\def\mn@eprint@dblp#1{\href {http://dblp.uni-trier.de/rec/bibtex/#1.xml}
  {dblp:#1}}
\def\mn@eprint@#1:#2:#3:#4\@nil{\def\@tempa {#1}\def\@tempb {#2}\def\@tempc
  {#3}\ifx \@tempc \@empty \let \@tempc \@tempb \let \@tempb \@tempa \fi \ifx
  \@tempb \@empty \def\@tempb {arXiv}\fi \@ifundefined
  {mn@eprint@\@tempb}{\@tempb:\@tempc}{\expandafter \expandafter \csname
  mn@eprint@\@tempb\endcsname \expandafter{\@tempc}}}

\bibitem[\protect\citeauthoryear{Abraham, Tanvir, Santiago, Ellis, Glazebrook
  \& Bergh}{Abraham et~al.}{1996}]{Abraham1996x}
Abraham R.~G.,  Tanvir N.~R.,  Santiago B.~X.,  Ellis R.~S.,  Glazebrook K.,
  Bergh S. v.~d.,  1996, \mn@doi [MNRAS] {10.1093/mnras/279.3.L47}, 279, L47

\bibitem[\protect\citeauthoryear{Abraham, van~den Bergh  \& Nair}{Abraham
  et~al.}{2003}]{Abraham2003}
Abraham R.~G.,  van~den Bergh S.,   Nair P.,  2003, \mn@doi [ApJ]
  {10.1086/373919}, 588, 218

\bibitem[\protect\citeauthoryear{Baes, Verstappen, Looze, Fritz, Saftly,
  P{\'e}rez, Stalevski  \& Valcke}{Baes et~al.}{2011}]{Baes2011}
Baes M.,  Verstappen J.,  Looze I.~D.,  Fritz J.,  Saftly W.,  P{\'e}rez E.~V.,
   Stalevski M.,   Valcke S.,  2011, \mn@doi [ApJS]
  {10.1088/0067-0049/196/2/22}, 196, 22

\bibitem[\protect\citeauthoryear{Baldry et~al.,}{Baldry
  et~al.}{2018}]{Baldry2018}
Baldry I.~K.,  et~al., 2018, \mn@doi [MNRAS] {10.1093/mnras/stx3042}, 474, 3875

\bibitem[\protect\citeauthoryear{Barbary}{Barbary}{2016}]{Barbary2016}
Barbary K.,  2016, \mn@doi [JOSS] {10.21105/joss.00058}, 1, 58

\bibitem[\protect\citeauthoryear{Bershady, Jangren  \& Conselice}{Bershady
  et~al.}{2000}]{Bershady2000}
Bershady M.~A.,  Jangren A.,   Conselice C.~J.,  2000, \mn@doi [AJ]
  {10.1086/301386}, 119, 2645

\bibitem[\protect\citeauthoryear{Bertin \& Arnouts}{Bertin \&
  Arnouts}{1996}]{Bertin1996}
Bertin E.,  Arnouts S.,  1996, \mn@doi [A\&ASS] {10.1051/aas:1996164}, 117, 393

\bibitem[\protect\citeauthoryear{Besla et~al.,}{Besla et~al.}{2018}]{Besla2018}
Besla G.,  et~al., 2018, \mn@doi [MNRAS] {10.1093/mnras/sty2041}, 480, 3376

\bibitem[\protect\citeauthoryear{Bignone, Tissera, Sillero, Pedrosa, Pellizza
  \& Lambas}{Bignone et~al.}{2016}]{Bignone2016}
Bignone L.~A.,  Tissera P.~B.,  Sillero E.,  Pedrosa S.~E.,  Pellizza L.~J.,
  Lambas D.~G.,  2016, \mn@doi [MNRAS] {10.1093/mnras/stw2788}, 465, 1106

\bibitem[\protect\citeauthoryear{Bottrell et~al.,}{Bottrell
  et~al.}{2019}]{Bottrell2019}
Bottrell C.,  et~al., 2019, \mn@doi [MNRAS] {10.1093/mnras/stz2934}, 490, 5390

\bibitem[\protect\citeauthoryear{Breiman}{Breiman}{2001}]{Breiman2001}
Breiman L.,  2001, \mn@doi [Mach. Learn.] {10.1023/A:1010933404324}, 45, 5

\bibitem[\protect\citeauthoryear{Bruzual \& Charlot}{Bruzual \&
  Charlot}{2003}]{Bruzual2003}
Bruzual G.,  Charlot S.,  2003, \mn@doi [MNRAS]
  {10.1046/j.1365-8711.2003.06897.x}, 344, 1000

\bibitem[\protect\citeauthoryear{Camps \& Baes}{Camps \&
  Baes}{2015}]{Camps2015}
Camps P.,  Baes M.,  2015, \mn@doi [Astron. Comput.]
  {10.1016/j.ascom.2014.10.004}, 9, 20

\bibitem[\protect\citeauthoryear{Casteels et~al.,}{Casteels
  et~al.}{2014}]{Casteels2014}
Casteels K. R.~V.,  et~al., 2014, \mn@doi [MNRAS] {10.1093/mnras/stu1799}, 445,
  1157

\bibitem[\protect\citeauthoryear{Chambers et~al.,}{Chambers
  et~al.}{2016}]{Chambers2016}
Chambers K.~C.,  et~al., 2016, The Pan-STARRS1 Surveys,
  \mn@doi{10.48550/ARXIV.1612.05560}, \url {https://arxiv.org/abs/1612.05560}

\bibitem[\protect\citeauthoryear{Chawla, Bowyer, Hall  \& Kegelmeyer}{Chawla
  et~al.}{2002}]{Chawla2002}
Chawla N.~V.,  Bowyer K.~W.,  Hall L.~O.,   Kegelmeyer W.~P.,  2002, \mn@doi
  [JAIR] {10.1613/jair.953}, 16, 321

\bibitem[\protect\citeauthoryear{Conselice}{Conselice}{2003}]{Conselice2003}
Conselice C.~J.,  2003, \mn@doi [ApJS] {10.1086/375001}, 147, 1

\bibitem[\protect\citeauthoryear{Conselice, Bershady  \& Jangren}{Conselice
  et~al.}{2000}]{Conselice2000}
Conselice C.~J.,  Bershady M.~A.,   Jangren A.,  2000, \mn@doi [ApJ]
  {10.1086/308300}, 529, 886

\bibitem[\protect\citeauthoryear{Crain et~al.,}{Crain et~al.}{2015}]{Crain2015}
Crain R.~A.,  et~al., 2015, \mn@doi [MNRAS] {10.1093/mnras/stv725}, 450, 1937

\bibitem[\protect\citeauthoryear{Davis, Efstathiou, Frenk  \& White}{Davis
  et~al.}{1985}]{Davis1985}
Davis M.,  Efstathiou G.,  Frenk C.~S.,   White S. D.~M.,  1985, \mn@doi [ApJ]
  {10.1086/163168}, 292, 371

\bibitem[\protect\citeauthoryear{Dolag, Borgani, Murante  \& Springel}{Dolag
  et~al.}{2009}]{Dolag2009}
Dolag K.,  Borgani S.,  Murante G.,   Springel V.,  2009, \mn@doi [MNRAS]
  {10.1111/j.1365-2966.2009.15034.x}, 399, 497

\bibitem[\protect\citeauthoryear{{Donnari} et~al.,}{{Donnari}
  et~al.}{2019}]{2019MNRAS.485.4817D}
{Donnari} M.,  et~al., 2019, \mn@doi [MNRAS] {10.1093/mnras/stz712}, \href
  {https://ui.adsabs.harvard.edu/abs/2019MNRAS.485.4817D} {485, 4817}

\bibitem[\protect\citeauthoryear{Dubois et~al.,}{Dubois
  et~al.}{2014}]{Dubois2014}
Dubois Y.,  et~al., 2014, \mn@doi [MNRAS] {10.1093/mnras/stu1227}, 444, 1453

\bibitem[\protect\citeauthoryear{Duncan et~al.,}{Duncan
  et~al.}{2019}]{Duncan2019}
Duncan K.,  et~al., 2019, \mn@doi [ApJ] {10.3847/1538-4357/ab148a}, 876, 110

\bibitem[\protect\citeauthoryear{Ellison, Patton, Mendel  \& Scudder}{Ellison
  et~al.}{2011}]{Ellison2011}
Ellison S.~L.,  Patton D.~R.,  Mendel J.~T.,   Scudder J.~M.,  2011, \mn@doi
  [MNRAS] {10.1111/j.1365-2966.2011.19624.x}, 418, 2043

\bibitem[\protect\citeauthoryear{Ellison, Viswanathan, Patton, Bottrell,
  McConnachie, Gwyn  \& Cuillandre}{Ellison et~al.}{2019}]{Ellison2019}
Ellison S.~L.,  Viswanathan A.,  Patton D.~R.,  Bottrell C.,  McConnachie
  A.~W.,  Gwyn S.,   Cuillandre J.-C.,  2019, \mn@doi [MNRAS]
  {10.1093/mnras/stz1431}, 487, 2491

\bibitem[\protect\citeauthoryear{Freeman, Izbicki, Lee, Newman, Conselice,
  Koekemoer, Lotz  \& Mozena}{Freeman et~al.}{2013}]{Freeman2013}
Freeman P.~E.,  Izbicki R.,  Lee A.~B.,  Newman J.~A.,  Conselice C.~J.,
  Koekemoer A.~M.,  Lotz J.~M.,   Mozena M.,  2013, \mn@doi [MNRAS]
  {10.1093/mnras/stt1016}, 434, 282

\bibitem[\protect\citeauthoryear{Genel et~al.,}{Genel et~al.}{2014}]{Genel2014}
Genel S.,  et~al., 2014, \mn@doi [MNRAS] {10.1093/mnras/stu1654}, 445, 175

\bibitem[\protect\citeauthoryear{Groves, Dopita, Sutherland, Kewley, Fischera,
  Leitherer, Brandl  \& van Breugel}{Groves et~al.}{2008}]{Groves2008}
Groves B.,  Dopita M.~A.,  Sutherland R.~S.,  Kewley L.~J.,  Fischera J.,
  Leitherer C.,  Brandl B.,   van Breugel W.,  2008, \mn@doi [ApJS]
  {10.1086/528711}, 176, 438

\bibitem[\protect\citeauthoryear{Guo \& White}{Guo \& White}{2008}]{Guo2008}
Guo Q.,  White S. D.~M.,  2008, \mn@doi [MNRAS]
  {10.1111/j.1365-2966.2007.12619.x}, 384, 2

\bibitem[\protect\citeauthoryear{Hernquist \& Katz}{Hernquist \&
  Katz}{1989}]{Hernquist1989}
Hernquist L.,  Katz N.,  1989, \mn@doi [ApJS] {10.1086/191344}, 70, 419

\bibitem[\protect\citeauthoryear{Hopkins et~al.,}{Hopkins
  et~al.}{2010}]{Hopkins2010}
Hopkins P.~F.,  et~al., 2010, \mn@doi [ApJ] {10.1088/0004-637X/715/1/202}, 715,
  202

\bibitem[\protect\citeauthoryear{Huertas-Company et~al.,}{Huertas-Company
  et~al.}{2015}]{HuertasCompany2015}
Huertas-Company M.,  et~al., 2015, \mn@doi [ApJS] {10.1088/0067-0049/221/1/8},
  221, 8

\bibitem[\protect\citeauthoryear{{Huertas-Company} et~al.,}{{Huertas-Company}
  et~al.}{2019}]{Huertas-Company2019}
{Huertas-Company} M.,  et~al., 2019, \mn@doi [MNRAS] {10.1093/mnras/stz2191},
  489, 1859

\bibitem[\protect\citeauthoryear{Kartaltepe et~al.,}{Kartaltepe
  et~al.}{2007}]{Kartaltepe2007}
Kartaltepe J.~S.,  et~al., 2007, \mn@doi [ApJS] {10.1086/519953}, 172, 320

\bibitem[\protect\citeauthoryear{Kewley, Geller  \& Barton}{Kewley
  et~al.}{2006}]{Kewley2006}
Kewley L.~J.,  Geller M.~J.,   Barton E.~J.,  2006, \mn@doi [AJ]
  {10.1086/500295}, 131, 2004

\bibitem[\protect\citeauthoryear{Kuijken et~al.,}{Kuijken
  et~al.}{2019}]{Kuijken2019}
Kuijken K.,  et~al., 2019, \mn@doi [A\&A] {10.1051/0004-6361/201834918}, 625,
  A2

\bibitem[\protect\citeauthoryear{Lema{\^i}tre, Nogueira  \&
  Aridas}{Lema{\^i}tre et~al.}{2017}]{Lemaitre2017}
Lema{\^i}tre G.,  Nogueira F.,   Aridas C.~K.,  2017, JMLR, 18, 1

\bibitem[\protect\citeauthoryear{Lin et~al.,}{Lin et~al.}{2004}]{Lin2004}
Lin L.,  et~al., 2004, \mn@doi [ApJ] {10.1086/427183}, 617, L9

\bibitem[\protect\citeauthoryear{Liske et~al.,}{Liske et~al.}{2015}]{Liske2015}
Liske J.,  et~al., 2015, \mn@doi [MNRAS] {10.1093/mnras/stv1436}, 452, 2087

\bibitem[\protect\citeauthoryear{Lotz, Primack  \& Madau}{Lotz
  et~al.}{2004}]{Lotz2004}
Lotz J.~M.,  Primack J.,   Madau P.,  2004, \mn@doi [AJ] {10.1086/421849}, 128,
  163

\bibitem[\protect\citeauthoryear{Lotz et~al.,}{Lotz et~al.}{2008}]{Lotz2008}
Lotz J.~M.,  et~al., 2008, \mn@doi [ApJ] {10.1086/523659}, 672, 177

\bibitem[\protect\citeauthoryear{Lotz, Jonsson, Cox, Croton, Primack,
  Somerville  \& Stewart}{Lotz et~al.}{2011}]{Lotz2011}
Lotz J.~M.,  Jonsson P.,  Cox T.~J.,  Croton D.,  Primack J.~R.,  Somerville
  R.~S.,   Stewart K.,  2011, \mn@doi [ApJ] {10.1088/0004-637X/742/2/103}, 742,
  103

\bibitem[\protect\citeauthoryear{Marinacci et~al.,}{Marinacci
  et~al.}{2018}]{Marinacci2018}
Marinacci F.,  et~al., 2018, \mn@doi [MNRAS] {10.1093/mnras/sty2206}, 480, 5113

\bibitem[\protect\citeauthoryear{{Martin} et~al.,}{{Martin}
  et~al.}{2021}]{2021MNRAS.500.4937M}
{Martin} G.,  et~al., 2021, \mn@doi [MNRAS] {10.1093/mnras/staa3443}, \href
  {https://ui.adsabs.harvard.edu/abs/2021MNRAS.500.4937M} {500, 4937}

\bibitem[\protect\citeauthoryear{Naiman et~al.,}{Naiman
  et~al.}{2018}]{Naiman2018}
Naiman J.~P.,  et~al., 2018, \mn@doi [MNRAS] {10.1093/mnras/sty618}, 477, 1206

\bibitem[\protect\citeauthoryear{Nelson et~al.,}{Nelson
  et~al.}{2018}]{Nelson2018}
Nelson D.,  et~al., 2018, \mn@doi [MNRAS] {10.1093/mnras/stx3040}, 475, 624

\bibitem[\protect\citeauthoryear{Nelson et~al.,}{Nelson
  et~al.}{2019a}]{Nelson2019}
Nelson D.,  et~al., 2019a, \mn@doi [Comput. Astrophys.]
  {10.1186/s40668-019-0028-x}, 6, 2

\bibitem[\protect\citeauthoryear{{Nelson} et~al.,}{{Nelson}
  et~al.}{2019b}]{2019MNRAS.490.3234N}
{Nelson} D.,  et~al., 2019b, \mn@doi [MNRAS] {10.1093/mnras/stz2306}, \href
  {https://ui.adsabs.harvard.edu/abs/2019MNRAS.490.3234N} {490, 3234}

\bibitem[\protect\citeauthoryear{Nelson et~al.,}{Nelson
  et~al.}{2021}]{10.1093/mnras/stab2131}
Nelson E.~J.,  et~al., 2021, \mn@doi [MNRAS] {10.1093/mnras/stab2131}, 508, 219

\bibitem[\protect\citeauthoryear{{Nevin}, {Blecha}, {Comerford}  \&
  {Greene}}{{Nevin} et~al.}{2019}]{2019ApJ...872...76N}
{Nevin} R.,  {Blecha} L.,  {Comerford} J.,   {Greene} J.,  2019, \mn@doi [ApJ]
  {10.3847/1538-4357/aafd34}, \href
  {https://ui.adsabs.harvard.edu/abs/2019ApJ...872...76N} {872, 76}

\bibitem[\protect\citeauthoryear{Patton, Torrey, Ellison, Mendel  \&
  Scudder}{Patton et~al.}{2013}]{Patton2013}
Patton D.~R.,  Torrey P.,  Ellison S.~L.,  Mendel J.~T.,   Scudder J.~M.,
  2013, \mn@doi [MNRAS Letters] {10.1093/mnrasl/slt058}, 433, L59

\bibitem[\protect\citeauthoryear{Pawlik, Wild, Walcher, Johansson, Villforth,
  Rowlands, {Mendez-Abreu}  \& Hewlett}{Pawlik et~al.}{2016}]{Pawlik2016}
Pawlik M.~M.,  Wild V.,  Walcher C.~J.,  Johansson P.~H.,  Villforth C.,
  Rowlands K.,  {Mendez-Abreu} J.,   Hewlett T.,  2016, \mn@doi [MNRAS]
  {10.1093/mnras/stv2878}, 456, 3032

\bibitem[\protect\citeauthoryear{Pearson et~al.,}{Pearson
  et~al.}{2019}]{Pearson2019}
Pearson W.~J.,  et~al., 2019, \mn@doi [A\&A] {10.1051/0004-6361/201936337},
  631, A51

\bibitem[\protect\citeauthoryear{Pedregosa et~al.,}{Pedregosa
  et~al.}{2011}]{Pedregosa2011}
Pedregosa F.,  et~al., 2011, JMLR, 12, 2825

\bibitem[\protect\citeauthoryear{{Pillepich} et~al.,}{{Pillepich}
  et~al.}{2018a}]{2018MNRAS.473.4077P}
{Pillepich} A.,  et~al., 2018a, \mn@doi [MNRAS] {10.1093/mnras/stx2656}, \href
  {https://ui.adsabs.harvard.edu/abs/2018MNRAS.473.4077P} {473, 4077}

\bibitem[\protect\citeauthoryear{Pillepich et~al.,}{Pillepich
  et~al.}{2018b}]{Pillepich2018}
Pillepich A.,  et~al., 2018b, \mn@doi [MNRAS] {10.1093/mnras/stx3112}, 475, 648

\bibitem[\protect\citeauthoryear{{Pillepich} et~al.,}{{Pillepich}
  et~al.}{2019}]{2019MNRAS.490.3196P}
{Pillepich} A.,  et~al., 2019, \mn@doi [MNRAS] {10.1093/mnras/stz2338}, \href
  {https://ui.adsabs.harvard.edu/abs/2019MNRAS.490.3196P} {490, 3196}

\bibitem[\protect\citeauthoryear{{Planck Collaboration XIII}}{{Planck
  Collaboration XIII}}{2016}]{Ade2016}
{Planck Collaboration XIII} 2016, \mn@doi [A\&A] {10.1051/0004-6361/201525830},
  594, A13

\bibitem[\protect\citeauthoryear{{Price-Whelan} et~al.,}{{Price-Whelan}
  et~al.}{2018}]{Price-Whelan2018}
{Price-Whelan} a. A.~M.,  et~al., 2018, \mn@doi [AJ]
  {10.3847/1538-3881/aabc4f}, 156, 123

\bibitem[\protect\citeauthoryear{Propris, Liske, Driver, Allen  \&
  Cross}{Propris et~al.}{2005}]{Propris2005}
Propris R.~D.,  Liske J.,  Driver S.~P.,  Allen P.~D.,   Cross N. J.~G.,  2005,
  \mn@doi [AJ] {10.1086/433169}, 130, 1516

\bibitem[\protect\citeauthoryear{Robitaille et~al.,}{Robitaille
  et~al.}{2013}]{Robitaille2013}
Robitaille T.~P.,  et~al., 2013, \mn@doi [A\&A] {10.1051/0004-6361/201322068},
  558, A33

\bibitem[\protect\citeauthoryear{{Rodriguez-Gomez} et~al.,}{{Rodriguez-Gomez}
  et~al.}{2015}]{Rodriguez-Gomez2015}
{Rodriguez-Gomez} V.,  et~al., 2015, \mn@doi [MNRAS] {10.1093/mnras/stv264},
  449, 49

\bibitem[\protect\citeauthoryear{{Rodriguez-Gomez} et~al.,}{{Rodriguez-Gomez}
  et~al.}{2019}]{Rodriguez-Gomez2019}
{Rodriguez-Gomez} V.,  et~al., 2019, \mn@doi [MNRAS] {10.1093/mnras/sty3345},
  483, 4140

\bibitem[\protect\citeauthoryear{{Rose} et~al.,}{{Rose}
  et~al.}{2022}]{2022arXiv220811164R}
{Rose} C.,  et~al., 2022, arXiv e-prints, \href
  {https://ui.adsabs.harvard.edu/abs/2022arXiv220811164R} {p. arXiv:2208.11164}

\bibitem[\protect\citeauthoryear{Schaye et~al.,}{Schaye
  et~al.}{2015}]{Schaye2015}
Schaye J.,  et~al., 2015, \mn@doi [MNRAS] {10.1093/mnras/stu2058}, 446, 521

\bibitem[\protect\citeauthoryear{Scudder, Ellison, Torrey, Patton  \&
  Mendel}{Scudder et~al.}{2012}]{Scudder2012}
Scudder J.~M.,  Ellison S.~L.,  Torrey P.,  Patton D.~R.,   Mendel J.~T.,
  2012, \mn@doi [MNRAS] {10.1111/j.1365-2966.2012.21749.x}, 426, 549

\bibitem[\protect\citeauthoryear{Sijacki, Vogelsberger, Genel, Springel,
  Torrey, Snyder, Nelson  \& Hernquist}{Sijacki et~al.}{2015}]{Sijacki2015}
Sijacki D.,  Vogelsberger M.,  Genel S.,  Springel V.,  Torrey P.,  Snyder
  G.~F.,  Nelson D.,   Hernquist L.,  2015, \mn@doi [MNRAS]
  {10.1093/mnras/stv1340}, 452, 575

\bibitem[\protect\citeauthoryear{Snyder, Lotz, Moody, Peth, Freeman, Ceverino,
  Primack  \& Dekel}{Snyder et~al.}{2015a}]{Snyder2015a}
Snyder G.~F.,  Lotz J.,  Moody C.,  Peth M.,  Freeman P.,  Ceverino D.,
  Primack J.,   Dekel A.,  2015a, \mn@doi [MNRAS] {10.1093/mnras/stv1231}, 451,
  4290

\bibitem[\protect\citeauthoryear{Snyder et~al.,}{Snyder
  et~al.}{2015b}]{Snyder2015b}
Snyder G.~F.,  et~al., 2015b, \mn@doi [MNRAS] {10.1093/mnras/stv2078}, 454,
  1886

\bibitem[\protect\citeauthoryear{Snyder, {Rodriguez-Gomez}, Lotz, Torrey,
  Quirk, Hernquist, Vogelsberger  \& Freeman}{Snyder et~al.}{2019}]{Snyder2019}
Snyder G.~F.,  {Rodriguez-Gomez} V.,  Lotz J.~M.,  Torrey P.,  Quirk A. C.~N.,
  Hernquist L.,  Vogelsberger M.,   Freeman P.~E.,  2019, \mn@doi [MNRAS]
  {10.1093/mnras/stz1059}, 486, 3702

\bibitem[\protect\citeauthoryear{Springel, Yoshida  \& White}{Springel
  et~al.}{2001a}]{Springel2001}
Springel V.,  Yoshida N.,   White S. D.~M.,  2001a, \mn@doi [New Astron.]
  {10.1016/S1384-1076(01)00042-2}, 6, 79

\bibitem[\protect\citeauthoryear{Springel, White, Tormen  \&
  Kauffmann}{Springel et~al.}{2001b}]{Springel2001a}
Springel V.,  White S. D.~M.,  Tormen G.,   Kauffmann G.,  2001b, \mn@doi
  [MNRAS] {10.1046/j.1365-8711.2001.04912.x}, 328, 726

\bibitem[\protect\citeauthoryear{Springel et~al.,}{Springel
  et~al.}{2018}]{Springel2018}
Springel V.,  et~al., 2018, \mn@doi [MNRAS] {10.1093/mnras/stx3304}, 475, 676

\bibitem[\protect\citeauthoryear{Stewart, Bullock, Barton  \& Wechsler}{Stewart
  et~al.}{2009}]{Stewart2009}
Stewart K.~R.,  Bullock J.~S.,  Barton E.~J.,   Wechsler R.~H.,  2009, \mn@doi
  [ApJ] {10.1088/0004-637X/702/2/1005}, 702, 1005

\bibitem[\protect\citeauthoryear{Taylor et~al.,}{Taylor
  et~al.}{2011}]{Taylor2011}
Taylor E.~N.,  et~al., 2011, \mn@doi [MNRAS]
  {10.1111/j.1365-2966.2011.19536.x}, 418, 1587

\bibitem[\protect\citeauthoryear{Thorp, Ellison, Simard, S{\'a}nchez  \&
  Antonio}{Thorp et~al.}{2019}]{Thorp2019}
Thorp M.~D.,  Ellison S.~L.,  Simard L.,  S{\'a}nchez S.~F.,   Antonio B.,
  2019, \mn@doi [MNRAS Letters] {10.1093/mnrasl/sly185}, 482, L55

\bibitem[\protect\citeauthoryear{Torrey et~al.,}{Torrey
  et~al.}{2015}]{Torrey2015}
Torrey P.,  et~al., 2015, \mn@doi [MNRAS] {10.1093/mnras/stu2592}, 447, 2753

\bibitem[\protect\citeauthoryear{{Torrey} et~al.,}{{Torrey}
  et~al.}{2019}]{2019MNRAS.484.5587T}
{Torrey} P.,  et~al., 2019, \mn@doi [MNRAS] {10.1093/mnras/stz243}, \href
  {https://ui.adsabs.harvard.edu/abs/2019MNRAS.484.5587T} {484, 5587}

\bibitem[\protect\citeauthoryear{Towns et~al.,}{Towns et~al.}{2014}]{Towns2014}
Towns J.,  et~al., 2014, \mn@doi [Comput. Sci. Eng.] {10.1109/MCSE.2014.80},
  16, 62

\bibitem[\protect\citeauthoryear{Trayford et~al.,}{Trayford
  et~al.}{2017}]{Trayford2017}
Trayford J.~W.,  et~al., 2017, \mn@doi [MNRAS] {10.1093/mnras/stx1051}, 470,
  771

\bibitem[\protect\citeauthoryear{Vogelsberger et~al.,}{Vogelsberger
  et~al.}{2014a}]{Vogelsberger2014}
Vogelsberger M.,  et~al., 2014a, \mn@doi [MNRAS] {10.1093/mnras/stu1536}, 444,
  1518

\bibitem[\protect\citeauthoryear{Vogelsberger et~al.,}{Vogelsberger
  et~al.}{2014b}]{Vogelsberger2014a}
Vogelsberger M.,  et~al., 2014b, \mn@doi [Nature] {10.1038/nature13316}, 509,
  177

\bibitem[\protect\citeauthoryear{{Wang}, {Pearson}  \&
  {Rodriguez-Gomez}}{{Wang} et~al.}{2020}]{2020A&A...644A..87W}
{Wang} L.,  {Pearson} W.~J.,   {Rodriguez-Gomez} V.,  2020, \mn@doi [A\&A]
  {10.1051/0004-6361/202038084}, \href
  {https://ui.adsabs.harvard.edu/abs/2020A&A...644A..87W} {644, A87}

\bibitem[\protect\citeauthoryear{Weinberger et~al.,}{Weinberger
  et~al.}{2017}]{Weinberger2017a}
Weinberger R.,  et~al., 2017, \mn@doi [MNRAS] {10.1093/mnras/stw2944}, 465,
  3291

\bibitem[\protect\citeauthoryear{Wen, Zheng  \& An}{Wen et~al.}{2014}]{Wen2014}
Wen Z.~Z.,  Zheng X.~Z.,   An F.~X.,  2014, \mn@doi [ApJ]
  {10.1088/0004-637X/787/2/130}, 787, 130

\bibitem[\protect\citeauthoryear{White \& Rees}{White \&
  Rees}{1978}]{White1978}
White S. D.~M.,  Rees M.~J.,  1978, \mn@doi [MNRAS] {10.1093/mnras/183.3.341},
  183, 341

\bibitem[\protect\citeauthoryear{{Whitney}, {Ferreira}, {Conselice}  \&
  {Duncan}}{{Whitney} et~al.}{2021}]{Whitney2021}
{Whitney} A.,  {Ferreira} L.,  {Conselice} C.~J.,   {Duncan} K.,  2021, \mn@doi
  [ApJ] {10.3847/1538-4357/ac1422}, \href
  {https://ui.adsabs.harvard.edu/abs/2021ApJ...919..139W} {919, 139}

\bibitem[\protect\citeauthoryear{{de Jong} et~al.,}{{de Jong}
  et~al.}{2013}]{deJong2013}
{de Jong} J.~T.~A.,  et~al., 2013, The Messenger, \href
  {https://ui.adsabs.harvard.edu/abs/2013Msngr.154...44D} {154, 44}

\bibitem[\protect\citeauthoryear{Ćiprijanović, Snyder, Nord  \&
  Peek}{Ćiprijanović et~al.}{2020}]{Ciprijanovic2020}
Ćiprijanović A.,  Snyder G.,  Nord B.,   Peek J.,  2020, \mn@doi [Astron.
  Comput.] {https://doi.org/10.1016/j.ascom.2020.100390}, 32, 100390

\makeatother
\end{thebibliography}
\end{document}